\documentclass[aps, a4paper,superscriptaddress, nofootinbib, twocolumn, 10pt]{revtex4}
\usepackage{amsmath,amssymb, bm}
\usepackage{dsfont}
\usepackage{epsfig}
\usepackage{graphicx}
\usepackage{slashed}
\usepackage{color}
\usepackage{subfigure} 
\usepackage{tikz}
\newcommand{\be}{\begin{equation}}
\newcommand{\ee}{\end{equation}}

\newcommand{\beq}{\begin{equation}}
\newcommand{\eeq}{\end{equation}}
\newcommand{\bea}{\begin{eqnarray}}
\newcommand{\eea}{\end{eqnarray}}
\newcommand{\ba}{\begin{align}}
\newcommand{\ea}{\end{align}}
\newcommand{\bfig}{\begin{figure}}
\newcommand{\efig}{\end{figure}}

\newcommand{\wh}{\widehat}
\newcommand{\nn}{\nonumber}

\newcommand{\ah}{\hat a}

\begin{document}
~\vspace{1cm}
\title{ Higher-order perturbative coefficients in QCD from series acceleration by conformal mappings}
\author{Irinel Caprini}
\affiliation{Horia Hulubei National Institute for Physics and Nuclear Engineering,
P.O.B. MG-6, 077125 Bucharest-Magurele, Romania}
 
\begin{abstract} The present calculations in perturbative QCD reach the order $\alpha_s^4$  for several correlators calculated to five loops, and the  huge computational difficulties make unlikely  the full six-loop calculation in the near future. This situation has practical consequences; in particular the treatment of the higher orders of the perturbation series for the current-current correlator of light quarks  is one of the main sources of errors in the extraction of the strong coupling from hadronic $\tau$ decays.  Several approximate estimates of the next coefficients  of the corresponding Adler function have been proposed, using various arguments.  In the present paper we exploit the analytic structure of the Adler function in the Borel plane,  which allows the definition of an improved perturbative expansion in powers of a conformal variable which maps the cut Borel plane onto the unit disk.   The new expansions converge in a larger domain of the Borel plane and, when reexpanded in powers of the strong coupling,  yield definite values for the higher perturbative coefficients. We apply the method to the Adler function in the $\overline{\rm MS}$ scheme and to a suitable weighted integral of this function in the complex $s$ plane, chosen such as to avoid model-dependent assumptions on analyticity. Our results  $c_{5,1}=287 \pm 40$, $c_{6,1}=2948 \pm 208$ and $c_{7,1}=(1.89 \pm 0.75)\times 10^4$, for the six-, seven- and eigth-loop coefficients, respectively,  agree with a recent determination from Pad\'e approximants applied to the perturbative expansion of the hadronic $\tau$ decay rate.
\end{abstract}
\maketitle
\vspace{0.5cm} 
\section{ Introduction}
The perturbative QCD expansion  of  the Adler function in the chiral limit is known  to five loops \cite{BCK08}, the same order to which the renormalization-group $\beta$ function has been calculated \cite{BCKbeta}. The increased precision of perturbative calculations is important for the extraction of the strong coupling $\alpha_s$ from hadronic $\tau$ decays. However, having in view the computational difficulties, a six-loop result is not foreseen in the near future. In the absence of exact calculations, approximate values for the next coefficients have been proposed, based on various arguments.   
 Such predictions have been made also for the coefficient of the $\alpha_s^4$ term in Refs. \cite{BCK02, KS95, Kataev:1995vh}, before the exact five-loop calculation was available.  However, they
 have not been confirmed in general by the exact result reported in \cite{BCK08}. After the appearance of this result, estimates of the  six-loop coefficient have been made in 
Ref. \cite{BCK08} from the principle of fastest apparent convergence (FAC) \cite{FAC}; in Ref. \cite{BeJa} from the convergence of the expansion of the $\tau$ hadronic decay rate;  and more recently, in Ref. \cite{Boito:2018rwt} from Pad\'e approximants applied to the perturbative expansion of the hadronic $\tau$ decay rate.

The prediction of the higher unknown terms in a series expansion from the knowledge of the first few terms may look like a miracle. Actually, this prediction would be impossible without the knowledge of some theoretical properties of the expanded function, available {\em a priori}, independent of the series expansion. In the present paper we exploit the analytic structure of the Adler function in the Borel plane,  which encodes the large-order behavior of the perturbative expansion. We then apply a procedure of series acceleration by conformal mappings,  proposed in \cite{CaFi1998} and investigated further in Refs. \cite{CaFi2000}-\cite{Caprini:2019osi}. Instead of the standard Taylor expansion of the Borel transform, we define new expansions, in powers of a conformal variable which maps the cut Borel plane onto the unit disk. These expansions converge in a larger domain of the Borel plane and have a better convergence rate.  Moreover, when reexpanded in powers of the strong coupling, they yield definite values for the higher-order perturbative coefficients. A prediction for the six-loop coefficient by this method was already reported in \cite{CaFi2009}. In the present work we perform a more systematic investigation of this problem.  

The paper is organized as follows. In the next section we briefly review the calculation of the Adler function and the hadronic width of the $\tau$ lepton  in perturbative QCD.  In  Sec. \ref{sec:conf} we introduce the modified perturbative expansions based on the conformal mapping of the Borel plane.
The prediction of the first three unknown coefficients from the expansion of the Adler function in the $\overline{\text{MS}}$ scheme  is investigated in Sec. \ref{sec:DMSbar}. We consider here also the prediction of the coefficients at large orders, using as a framework two renormalon-based models for the Adler function, reviewed for completeness in the Appendix.  In Sec. \ref{sec:DC} we briefly discuss the prediction of the six-loop coefficient in an alternative renormalization scheme, known as the $C$ scheme. In Sec. \ref{sec:delta0} we explore the possibility of extracting the next coefficients using the perturbative expansion of the hadronic decay width of the $\tau$ lepton, which is expressed as a weighted integral of the Adler function along a contour in the complex energy plane. In Sec. \ref{sec:weights} we discuss other weighted integrals and define a criterion for the choice of an optimal weight, which avoids model-dependent assumptions on the properties in the Borel plane. A suitable weight meeting this criterion is considered in Sec. \ref{sec:Ioptim} for the extraction of the  unknown perturbative coefficients of interest. In Sec. \ref{sec:final} we present the final results, obtained by averaging the results independent of {\em ad hoc} assumptions obtained in our analysis.  The last section contains a summary and our conclusions. 
\section{Adler function and $\tau$ hadronic width in perturbative QCD}\label{sec:Adler}
  We recall that the Adler function is the logarithmic derivative of the invariant amplitude of the two-current correlation tensor, $D(s)=-s \,d\Pi(s)/ds$, where $s$ is the momentum squared. We shall consider the reduced function $\widehat D(s)$ defined as:
\beq\label{eq:D}
\widehat{D}(s) \equiv 4 \pi^2 D(s) -1.
\eeq
From general principles of field theory, it is known that $\wh D(s)$ is an analytic function of real type [i.e. it satisfies the  Schwarz reflection property $\wh D(s^*)=\wh D^*(s)$] in the complex $s$ plane cut along the timelike axis for $s\ge 4 m_\pi^2$.

 At large spacelike momenta $s<0$, this function is given by the QCD perturbative expansion  
\beq\label{eq:hatD}
\widehat{D}(s) =\sum\limits_{n\ge 1} a_\mu^n \,
\sum\limits_{k=1}^{n} k\, c_{n,k}\, (\ln (-s/\mu^2))^{k-1},
\eeq
where $a_\mu \equiv \alpha_s(\mu^2)/\pi$ is the renormalized strong coupling in a certain renormalization scheme (RS) at an arbitrary scale $\mu$.
The  coefficients  $c_{n,1}$ are obtained from the calculation of  Feynman diagrams, while  $c_{n,k}$ with $k>1$ are obtained in terms of  $c_{m,1}$ with $m< n$  and the coefficients $\beta_n$ of the $\beta$ function, which governs the variation of the QCD coupling with the scale  in each RS:
 \begin{equation}\label{eq:rge}
 -\mu\frac {d a_\mu}
{d\mu}\equiv \beta(a_\mu)=\sum_{n\ge 1}
\beta_n a_\mu^{n+1}\,. \end{equation}
We recall that in mass-independent renormalization schemes the first two coefficients  $\beta_1$ and $\beta_2$ are scheme invariant, depending only on the number $n_f$ of active flavors, while  $\beta_n$ for $n\ge 3$ depend on the renormalization scheme. 
In the $\overline{{\rm MS}}$ scheme, the known coefficients  for $n_f=3$ are (cf. \cite{BCKbeta} and references therein)
\be\label{eq:betaj}
\beta_1=\frac{9}{2},~  \beta_2=8, ~ \beta_3= 20.12,~ \beta_4=94.46, ~\beta_5=268.16.
\ee

By choosing in (\ref{eq:hatD}) the scale $\mu^2=-s$, one  obtains the renormalization-group improved expansion
\beq\label{eq:hatD1}
\widehat{D}(s) =\sum\limits_{n\ge 1} c_{n,1}\, [a(-s)]^n,
\eeq
where $a(-s)\equiv \alpha_s(-s)/\pi$ is the running coupling.

The Adler function  was calculated in $\overline{{\rm MS}}$ to order $\alpha_s^4$  (see \cite{BCK08} and references therein). For $n_f=3$, the leading coefficients $c_{n,1}$  have the values
\be\label{eq:cn1}
c_{1,1}=1,\,\, c_{2,1}=1.640,\,\, c_{3,1}=6.371,\,\, c_{4,1}=49.076.
\ee
On the other hand, at large orders $n$ the coefficients $c_{n,1}$ are known to increase like $n!$  \cite{tHooft, Mueller1985,  Mueller1992,  Beneke}. Thus, the series (\ref{eq:hatD})  has a zero radius of convergence and is interpreted as an asymptotic expansion of $\widehat{D}(s)$ for $a_\mu\to 0$. As it is known, in some definite cases the expanded functions can be recovered from their divergent expansions through Borel summation. In the case of the Adler function,  the Borel transform  is defined by the power series
\be\label{eq:B}
 B_D(u)= \sum_{n=0}^\infty  b_n\, u^n,
\ee
where the coefficients $b_n$ are related to the perturbative coefficients $c_{n,1}$ by 
\be\label{eq:bn}
 b_n= \frac{c_{n+1,1}}{\beta_0^n \,n!}\,.
\ee
Here we used the standard notation $\beta_0=\beta_1/2$.

\begin{figure*}\vspace{-1cm}
\includegraphics[scale=0.18]{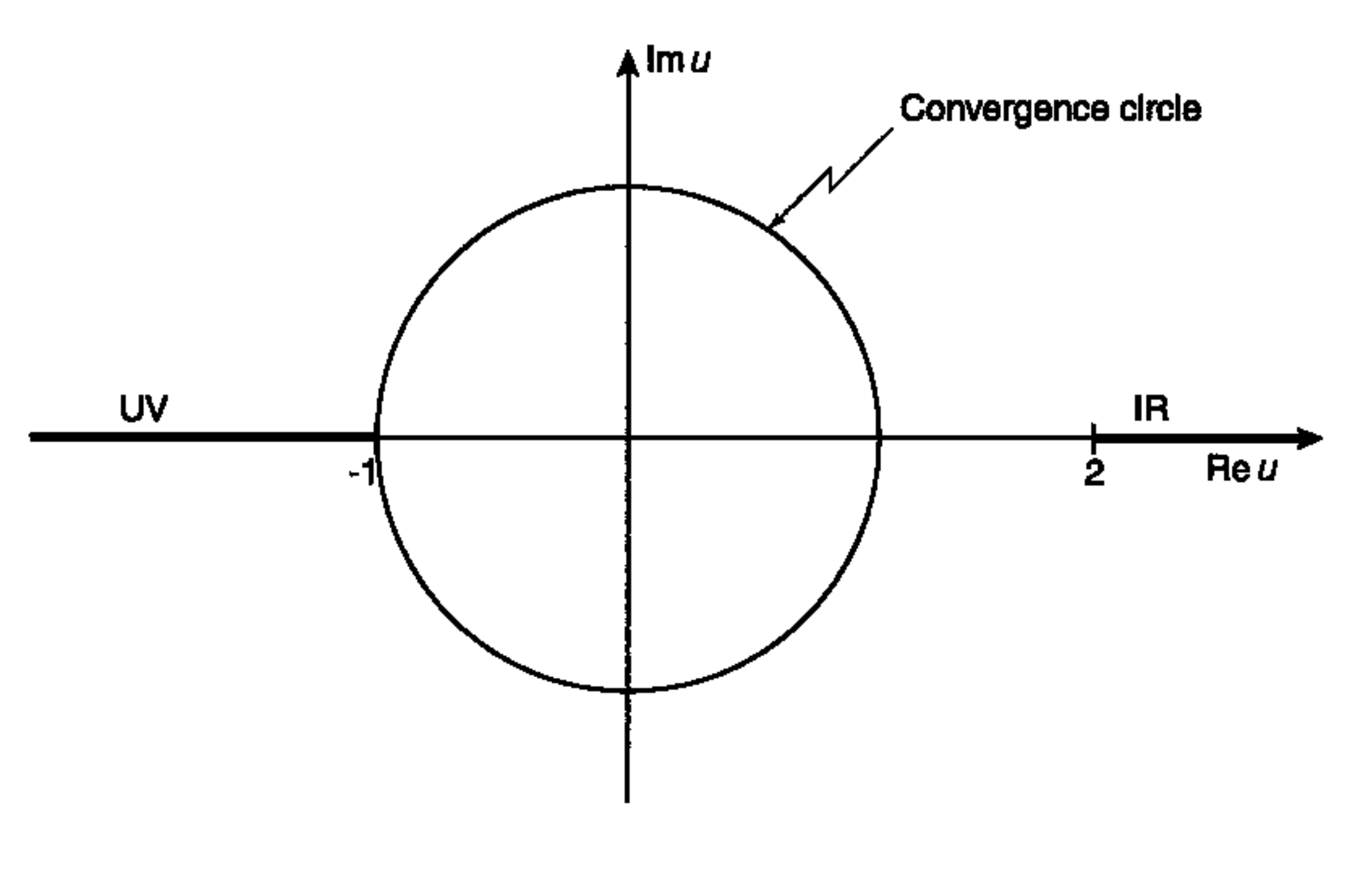}\hspace{0.3cm}\includegraphics[scale=0.27]{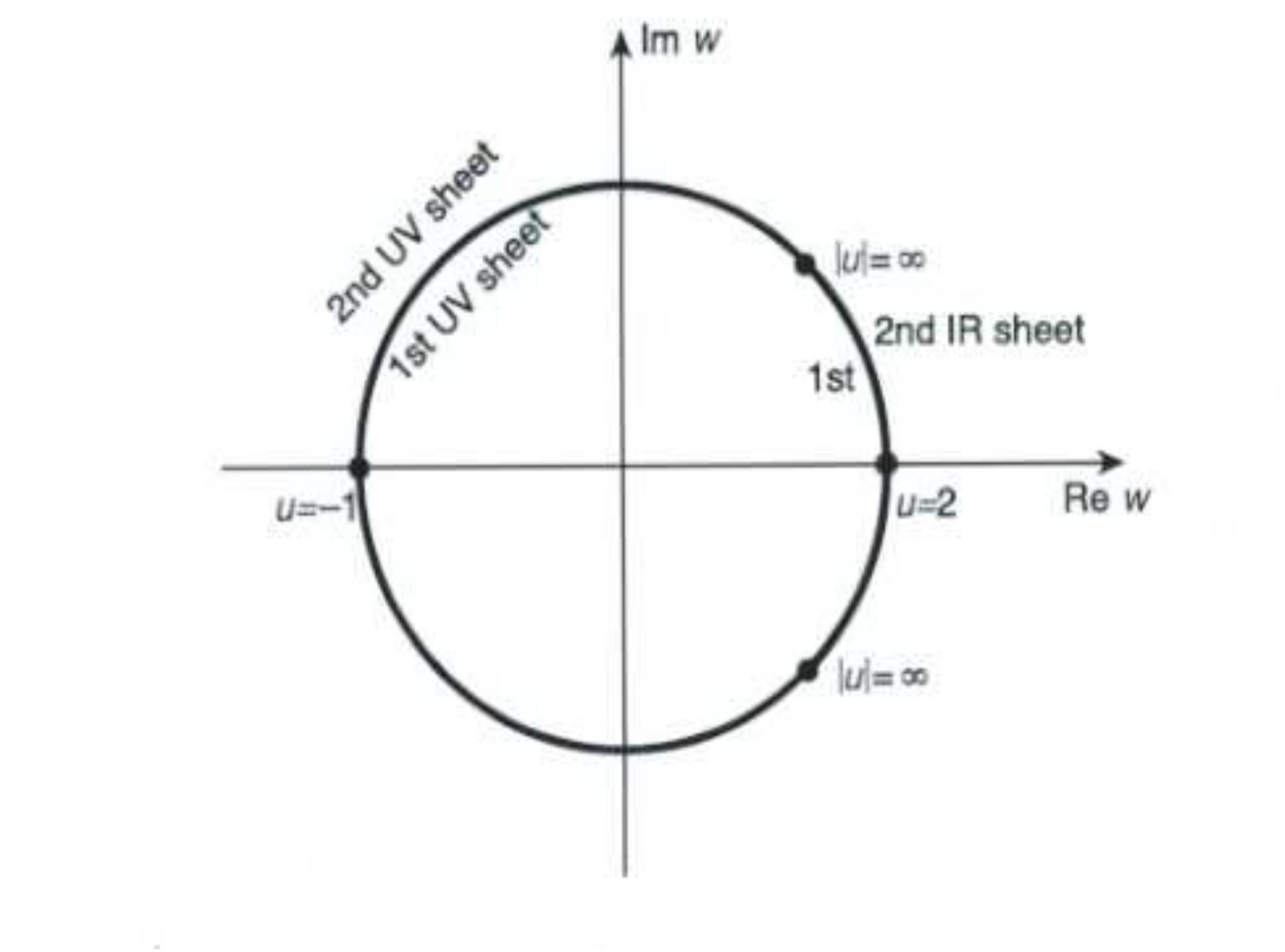}\hspace{0.5cm}\includegraphics[scale=0.119]{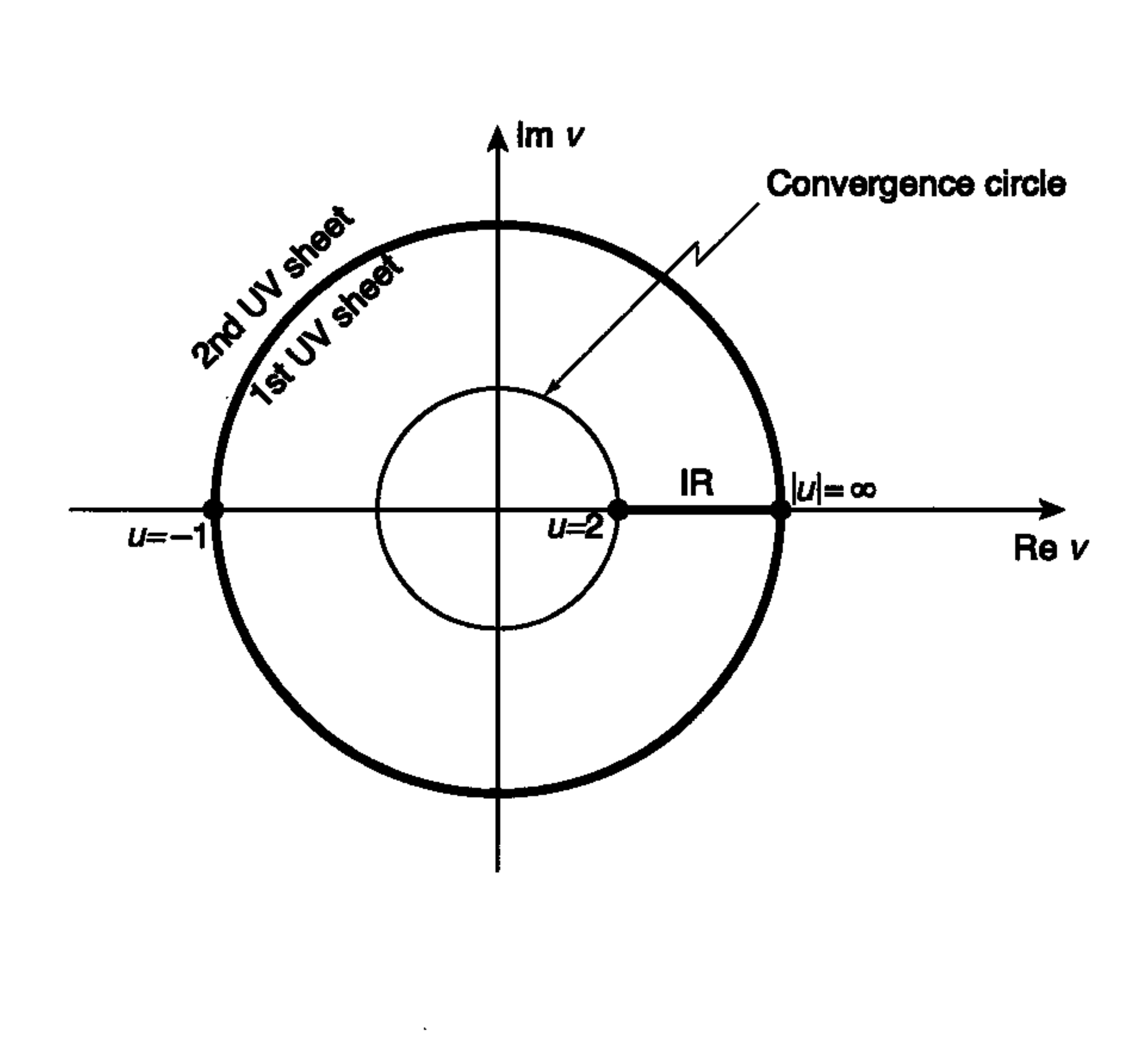}
\caption{Left: Borel plane of the Adler function. The circle indicates the convergence domain of the series (\ref{eq:B}). Middle: The  $w$ plane obtained by the conformal mapping (\ref{eq:w}). The IR and UV renormalons are mapped on the boundary of the unit disk.  Right: The  $v$ plane obtained by the conformal mapping (\ref{eq:v}). The cut $u\leq -1$ is mapped onto the unit circle. The cut $u\ge 2$ is mapped on a real segment inside the circle.
\label{fig:confBorel}}
\end{figure*}

The large-order increase of the coefficients of the perturbation series is encoded  in the singularities of the Borel transform in the complex $u$ plane. In the particular case of the Adler function,  $B_D(u)$ has singularities on the semiaxis $u\ge 2$, denoted as  infrared (IR) renormalons, and for $u\le -1$, denoted as ultraviolet (UV) renormalons (see Fig. \ref{fig:confBorel}, left panel). The names indicate the regions in the Feynman integrals responsible for the appearance of the corresponding singularities.  Moreover, the nature of the first branch points is known: near $u=-1$ and $u=2$,  $B_D(u)$ behaves like
\begin{equation}\label{eq:gammapowers}
 B_D(u) \sim \frac{r_1}{(1+u)^{\gamma_{1}}} \quad{\rm and} \quad  B_D(u)  \sim \frac{r_2}{(1-u/2)^{\gamma_{2}}}, 
\end{equation}
respectively, where the residues $r_1$ and $r_2$ are not known, but the exponents
$\gamma_1$ and $\gamma_2$ have been calculated in \cite{Mueller1985, Mueller1992, BBK, BeJa}. They are expressed in terms of the first coefficients $\beta_1$ and $\beta_2$ of the $\beta$ function, and for $n_f=3$  their values  are  
\begin{equation}\label{eq:gamma12}
\gamma_1 = 1.21,    \quad\quad   \gamma_2 = 2.58 \,. 
\end{equation}
  Apart from  the two cuts along the lines $u\geq 2$ and $u\leq -1$, it is assumed that no other singularities are  present in the complex $u$ plane \cite{Mueller1985}.

From the definition (\ref{eq:B}), it follows that the function $\wh D(s)$ defined by (\ref{eq:hatD1}) can be recovered formally from the Borel transform by the Laplace-Borel integral representation 
\be\label{eq:Laplace}
\wh D(s)=\frac{1}{\beta_0} \,\int\limits_0^\infty  
\exp{\left(\frac{-u}{\beta_0 a(-s)}\right)} \,  B_D(u)\, d u\,.
\ee
Actually, due to the singularities of $ B_D(u)$ for $u\ge 2$, the  integral (\ref{eq:Laplace}) is not defined and requires a regularization.  However,  this ambiguity will not affect our analysis, which will be restricted to the expansions of the Borel transform.

 We shall consider also the perturbative expansion of the $\tau$ hadronic width. The ratio $R_\tau$ of the total $\tau$ hadronic branching fraction to the electron branching fraction can be expressed as \cite{BeJa}
\be\label{eq:Rtau}
R_\tau= 3 \,S_{\rm EW} (|V_{ud} |^2 + |V_{us}|^2 ) (1 + \delta^{(0)} + \ldots),
\ee
where $S_{\rm EW}$ is an electroweak correction, $V_{ud}$ and $V_{us}$ are Cabibbo-Kobayashi-Maskawa  matrix elements, and $\delta^{(0)}$ is the perturbative QCD contribution. As shown in   \cite{Narison, Braaten, BrNaPi, Diberder}, $\delta^{(0)}$ can be expressed, using analyticity, by a weighted integral of the Adler function along a contour in the complex $s$ plane. In our normalization, this relation is \cite{BeJa}: 
\be\label{eq:delta0}
\delta^{(0)} =  \frac{1}{2\pi i} \oint\limits_{|s|=m_\tau^2}\, \frac{d s}{s} \left(1-\frac{s}{m_\tau^2}\right)^3\,\left(1+\frac{s}{m_\tau^2}\right) \wh D(s).
\ee

By inserting in the integral (\ref{eq:delta0}) the expansion   (\ref{eq:hatD}) at the fixed scale  $\mu=m_\tau$ and performing the integration with respect to $s$ of the coefficients, one obtains the fixed-order (FO) perturbative expansion\footnote{In the alternative prescription known as ``contour improved'' (CI), one inserts in  (\ref{eq:delta0})  the renormalization-group improved expansion (\ref{eq:hatD1}),  the running coupling $a(-s)$ being calculated by the numerical integration of the renormalization-group equation (\ref{eq:rge}) along the circle $|s|=m_\tau^2$.} of $\delta^{(0)}$
\beq\label{eq:del0gen}
\delta_{\rm FO}^{(0)} =\sum_{n\ge 1} d_n a_{\mu}^n, \quad \quad \mu=m_\tau.
\eeq

The first coefficients of this expansion read 
\bea\label{eq:dn}
&&d_1=1, \, d_2= 5.20, \,d_3=26.37, \,d_4=127.08, \nn\\
&& d_5= 307.8 + c_{5,1},\, d_6=-5848.2 + 17.81 c_{5,1} + c_{6,1},\nn\\ 
&&d_7=-97769.1 + 61.33\, c_{5,1} + 21.38\, c_{6,1} + c_{7,1},
\eea
where we have used the values given in  (\ref{eq:cn1}) and left free  the next coefficients $c_{n,1}$. 

In analogy with the expansion (\ref{eq:B}), we define the Borel transform $B_{\delta}(u)$ associated to the series (\ref{eq:del0gen}) by the Taylor expansion
\beq\label{eq:Bdelta}
B_{\delta}(u)= \sum_{n=0}^\infty  b'_n\, u^n,
\eeq
where the coefficients are defined as
\beq\label{eq:bnprime}
  b'_n= \frac{d_{n+1}}{\beta_0^n \,n!}\,.
\eeq
Then $\delta_{\rm FO}^{(0)}$ can be recovered  by the formal Laplace-Borel integral
\be\label{eq:Laplacedelta0}
\delta_{\rm FO}^{(0)}=\frac{1}{\beta_0} \,\int\limits_0^\infty  
\exp{\left(\frac{-u}{\beta_0 a_\mu}\right)} \,  B_{\delta}(u)\, d u\,, \quad \mu=m_\tau.
\ee
The analytic properties of the function  $B_{\delta}(u)$, which are important for the problem investigated in this work,  will be discussed in Sec. \ref{sec:delta0}.

Up to now we considered the perturbative expansions in the $\overline{\text{MS}}$ renormalization scheme. Recently, a different scheme was proposed in \cite{Cscheme, Cscheme1, Cscheme2} and was investigated further in \cite{Caprini:2018agy} by means of the conformal mapping approach to be presented in the next section. In this scheme, the coupling  $\hat a_\mu$ satisfies the renormalization-group equation 
\be\label{eq:betahat}
-\,\mu\,\frac{d \ah_\mu}{d\mu} \,\equiv\, \hat\beta(\ah_\mu) \,=\,
\frac{\beta_1 \ah_\mu^2}{\left(1 - \frac{\beta_2}{\beta_1}\, \ah_\mu\right)},
\ee  
which involves only the scheme-independent coefficients $\beta_1$ and $\beta_2$.  Furthermore, as shown in \cite{Cscheme1, Cscheme2}, the coupling $\ah_\mu$ depends on a single parameter, denoted as $C$, and the dependence of $\ah_\mu$ on this parameter is governed by the same scheme-independent function $\hat\beta$. 
 
The connection between the $C$-coupling $\ah_\mu$ and  the coupling $a_\mu$ in the  $\overline{{\rm MS}}$ renormalization scheme can be found by solving numerically a nonlinear equation given in  \cite{Cscheme}. From this equation 
one can obtain also the perturbative relations between the couplings $\ah_\mu$  and $a_\mu$:
\be\label{eq:atoahat} 
\ah_\mu(a_\mu)= \sum_{n\ge 1} \xi_n(C)\, a_\mu^n,\quad a_\mu(\ah_\mu)= \sum_{n\ge 1}\bar\xi_n(C)\, \ah_\mu^n.
\ee 
The explicit forms of these expansions are given in  Eqs. (7) and (8) of \cite{Cscheme}. From the comparison with the full solution, found numerically in \cite{Cscheme, Caprini:2018agy},   one can establish the range of $C$ where the perturbative expansions (\ref{eq:atoahat}) are valid. 

The expansion of the Adler function in powers of the $C$-coupling,  derived in a straightforward way using the perturbative relations (\ref{eq:atoahat}), has the generic form
\beq\label{eq:hatDC}
 \widehat{D}(s) =\sum\limits_{n\ge 1} \hat c_{n,1}(C)\, \hat a_\mu^n,
\eeq
For completeness, we write down the terms up to $n\leq 5$:
\bea\label{eq:hatDC1}
&&\hspace{-0.2cm}\wh D(s)= \ah_\mu + (1.64 + 2.25\, C)\,\ah_\mu^2  \nn\\
 && \hspace{-0.7cm}+\, (7.68 + 11.38\,C + 5.06 \,C^2)\,\ah_\mu^3 +\,(61.06 + 72.08 \,C  \nn \\
&& \hspace{-0.7cm}  + \, 47.40\, C^2 + 11.4\, C^3)\,\ah_\mu^4 + (c_{5,1}+ 65.5+ 677.7\, C \nn\\
&& \hspace{-0.7cm} +\,  408.6\, C^2 + 162.5\, C^3 + 25.6\, C^4) \ah_\mu^5+\ldots\,,  
\eea
where in the first four terms we have used as input the values (\ref{eq:cn1}). 

 Starting from (\ref{eq:hatDC}),  the Borel transform of the Adler function in the $C$ scheme is defined by the series
\be\label{eq:BhatC}
\widehat B(u, C)= \sum_{n=0}^\infty \hat b_n(C) u^n,
\eeq
where
\beq\label{eq:hatbn}
 \hat b_n(C)= \frac{\hat c_{n+1,1}(C)}{\beta_0^n \,n!}\,.
\ee
 
The Borel transform (\ref{eq:BhatC})  was introduced in \cite{Caprini:2018agy}, where it was used for the calculation of the Adler function and the $\tau$ hadronic width. In Sec. \ref{sec:DC} we will discuss the usefulness of the $C$ renormalization scheme for the prediction of the higher-order perturbative coefficients.

\section{\bf Series acceleration by conformal mappings}\label{sec:conf}

The singularities of $B_D(u)$ set a limitation on the convergence region of the power expansion (\ref{eq:B}): this series converges only inside  the circle $|u|=1$,  which passes through the first UV renormalon, shown in the left panel of Fig. \ref{fig:confBorel}. 
As it is known, the domain of convergence of a power series in a complex plane can be increased by expanding the function in powers of another variable,  which  performs the conformal mapping of the original plane  (or a part of it) onto a disk. 

The method of conformal mappings  was  introduced  in particle physics in  \cite{CiFi, Frazer, CCF} for improving the convergence of  the expansions of scattering amplitudes in powers of various kinematical variables. By expanding the amplitude in powers of the function that maps the original analyticity domain onto a unit disk, the new series converges in a larger region,  well beyond the convergence domain of the original expansion, and moreover has an increased  asymptotic convergence rate at points lying inside this domain. An important result proved in  \cite{CiFi, CCF}  is that the asymptotic convergence rate is maximal if the new variable maps the entire holomorphy domain of the expanded function onto the unit disk (a detailed proof is given in \cite{CaFi2011, Caprini:2019osi}).
This particular variable is known in the literature as the ``optimal conformal mapping''. 

In QCD, since the correlators are singular 
at the origin of the coupling plane \cite{tHooft},  the method of conformal mapping  is not applicable to the 
formal perturbative series in powers of the coupling\footnote{The conformal mapping of the coupling complex plane was used  in \cite{ZJ, ZJ1} by assuming  that the singularity is  shifted away from the origin by a certain amount at each finite perturbative order, and tends to the origin only for an infinite number of terms. The corresponding conformal mappings are known ``order-dependent'' mappings.}.  However, the method can be applied in a straightforward way to the Borel transform $B_D(u)$, which is holomorphic in a region containing the origin  $u = 0$  of the Borel complex plane and  can be expanded in powers of  the Borel variable as in (\ref{eq:B}).

 As shown for the first time in \cite{CaFi1998}, the optimal mapping, which ensures 
the  convergence of the corresponding power series in the entire doubly cut Borel plane,  is given by the function 
\begin{equation}\label{eq:w}
\tilde w(u)=\frac{\sqrt{1+u}-\sqrt{1-u/2}}{ \sqrt{1+u}+\sqrt{1-u/2}},
\end{equation}
whose inverse reads
\beq\label{eq:uw}
\tilde u(w)=\frac{8 w}{3-2 w+3 w^2}.
\eeq
One can check that the function $\tilde w(u)$  maps the complex  $u$ plane cut along the real axis for $u\ge 2$ and $u\le -1$ onto the interior
of the circle $\vert w\vert\, =\, 1$ in the complex plane $w\equiv \tilde w(u)$,  such that  the origin $u=0$ of the $u$ plane
corresponds to the origin $w=0$ of the $w$ plane, and the upper (lower) edges of the cuts are mapped onto the upper
(lower) semicircles in the  $w$ plane. 
By the  mapping (\ref{eq:w}), all  the singularities of the Borel transform, the  UV and IR  renormalons, are pushed on the boundary of the unit disk in the $w$  plane, all at equal distance from the origin (see the middle panel of Fig. \ref{fig:confBorel}). Therefore, the expansion of $B_D(u)$ as 
\be\label{eq:Bw}
B_D(u)=\sum_{n\ge 0} c_n \,w^n, \quad\quad w = \tilde w(u),
\ee
converges in the whole $u$ complex plane up to the cuts, i.e. in a much larger domain than the original series (\ref{eq:B}). According to the results mentioned above (proved in Ref. \cite{CaFi2011}), this expansion has the best asymptotic convergence rate compared to other expansions based on alternative conformal mappings.

The expansion can be further improved by exploiting  the fact that the  nature of the leading singularities of $B_D(u)$ in the Borel plane is known. Using (\ref{eq:w}), it is easy to check that 
\bea
(1+u)^{\gamma_1}&\sim & (1+w)^{2\gamma_1}, \quad \text{for}\, u\sim -1\nonumber\\
(1-u/2)^{\gamma_2}&\sim &  (1-w)^{2\gamma_2}, \quad \text{for}\, u\sim 2.
\eea
It follows that the product $B_D(u) (1+w)^{2\gamma_{1}} (1-w)^{2\gamma_{2}}$ will be finite at $u=-1$ and $u=2$. However, this product  still has singularities (branch points)  at  $u=-1$ and $u=2$, generated by the terms of $B_D(u)$ which are holomorphic at these points. Therefore, the optimal variable for the expansion of the product is the conformal mapping (\ref{eq:w}), which accounts for these singularities.  Using this remark, we shall adopt the expansion 
\be\label{eq:Bw1}
 B_D(u)=\frac{1}{(1+w)^{2\gamma_{1}} (1-w)^{2\gamma_{2}}} \sum_{n\ge 0} f_n\, w^n,
\ee
 proposed in \cite{CaFi2009} and investigated further in  \cite{CaFi2011}.

We note that a nonoptimal conformal mapping of the Borel plane, which takes into account only the position of the nearest singularity of $B_D(u)$, was suggested in  \cite{Mueller1992} and was used further in  \cite{Alta, SoSu}
 in order to reduce  the ambiguities  due to the UV renormalons. This mapping reads
\begin{equation}\label{eq:v}
\tilde v(u)=\frac{\sqrt{1+u}-1}{ \sqrt{1+u}+1},
\end{equation}
and has the inverse
\beq\label{eq:uv}
\tilde u(v)=\frac{4 v}{(1-v)^2}. 
\eeq
 The variable (\ref{eq:v}) maps the $u$ complex plane cut along the line $u\leq -1$  onto the unit disk $|v|<1$ in the plane  $v\equiv \tilde v(u)$, such that the point $u=-1$ becomes $v=-1$ and the point at infinity becomes $v=1$ (see the right panel of Fig. \ref{fig:confBorel}). In the $v$ plane, the image of the IR cut is the real segment $(\tilde v(2),\, 1)$ situated inside the circle. The expansion of the Borel function in this variable is
\beq\label{eq:Bv}
B_D(u)=\sum_{n\ge 0} g_n\, v^n.
\eeq
We can implement the nature of the first singularities expressed by (\ref{eq:gammapowers}) also in this variable.
Using the relations
\bea
(1+u)^{\gamma_1}&\sim & (1+v)^{2\gamma_1},\quad \text{for}\, u\sim -1\nonumber\\
 (1-u/2)^{\gamma_2}&\sim & (1-v/\tilde v(2))^{\gamma_2},\quad \text{for}\, u\sim 2,
\eea
which can be derived from (\ref{eq:v}),  we conclude that the product  $B_D(u) (1+v)^{2\gamma_{1}} (1-v/\tilde v(2))^{\gamma_{2}}$ is finite at $u=-1$ and $u=2$. By expanding this product in powers of $v$, we write the expansion of $B_D(u)$  as
\be\label{eq:Bv1}
 B_D(u)=\frac{1}{(1+v)^{2\gamma_{1}} (1-v/\tilde v(2))^{\gamma_{2}}} \sum_{n\ge 0} h_n\, v^n.
\ee
 We emphasize that in the expansions (\ref{eq:Bw1}) and  (\ref{eq:Bv1}), the global prefactors which implement the known behavior (\ref{eq:gammapowers}) near the first singularities are expressed in terms of the variable used in the power expansion.

\begin{table*}[htb]
\caption{Columns 2 to 4: Coefficient $c_{N,1}$ obtained from the knowledge of the coefficients $c_{n,1}$ for $n\leq N-1$  of the model  \cite{BeJa}, using the expansions (\ref{eq:Bv}) and  (\ref{eq:Bv1}) in powers of the nonoptimal mapping (\ref{eq:v}), and the expansions  (\ref{eq:Bw}) and  (\ref{eq:Bw1}) in powers of the optimal mapping (\ref{eq:w}). Last column: The exact perturbative coefficients $c_{N,1}$ of the model  \cite{BeJa}. } \vspace{0.1cm}
\label{tab:1}
 \renewcommand{\tabcolsep}{0.55pc} 
\renewcommand{\arraystretch}{1.15} 
\begin{tabular}{l|l l  |l l  | l}\hline\hline
$N$ \,\,\,    & Eq. (\ref{eq:Bv}) &Eq. (\ref{eq:Bw}) & Eq. (\ref{eq:Bv1})  & Eq. (\ref{eq:Bw1})& Exact $c_{N,1}$    \\\hline
4 & $-52.34$ &$ -17.61$ & 14.77 & 17.85 & 49.076 \\
5 &$ -932.45 $& $-270.46$ & 255.98& 255.73 & 283. \\
6 & $-14348.46$ &$-2290.94 $& 3096.35 & 2928.76 &  3275.45\\
7 & $-274384.$ &$-39054.7$ & 15740.1&  16308.73&  18758.4 \\
8 & $-5.12\times 10^6$ &$-272605.1$ & 350336.4 & 381151.6 & 388445.6\\
9 & $-1.14\times 10^8$ &$-6.89 \times 10^6$ & 455072.1 & 963059.1&  919119.2\\
10 & $-2.56 \times 10^9$& $-1.424 \times 10^7$ & $7.82 \times 10^7$ & $ 8.49 \times 10^7$&   $8.37 \times 10^7$\\
11 & $-6.68 \times 10^{10}$ & $-1.78 \times 10^9$ & $-5.74 \times 10^8$ & $-5.04 \times 10^8$& $-5.19 \times 10^8$\\
12 &$-1.76 \times 10^{12}$& $1.66 \times 10^{10}$ &  $3.36 \times 10^{10}$& $3.39 \times 10^{10}$ &  $3.38 \times 10^{10}$\\
13 & $-5.29 \times 10^{13}$ & $-8.47 \times 10^{11}$ & $-5.89 \times 10^{11}$ & $-6.04 \times 10^{11}$ & $-6.04 \times 10^{11}$\\
14 &$-1.61 \times 10^{15}$ & $1.98 \times 10^{13}$ &  $2.42 \times 10^{13}$ & $2.34 \times 10^{13}$& $ 2.34 \times 10^{13}$\\
15 & $-5.48 \times 10^{16}$ & $-7.09 \times 10^{14}$ & $-6.24 \times 10^{14}$ & $-6.53 \times 10^{14}$& $-6.52 \times 10^{14}$\\
16 & $-1.89 \times 10^{18}$ & $2.32 \times 10^{16}$&  $2.52 \times 10^{16}$& $2.42 \times 10^{16}$ &  $2.42 \times 10^{16}$\\
17 & $-7.22 \times 10^{19}$&$-8.62 \times 10^{17}$& $-8.12 \times 10^{17}$&  $-8.46 \times 10^{17}$& $-8.46 \times 10^{17}$\\
18 & $-2.78 \times 10^{21}$ & $3.33 \times 10^{19} $&  $ 3.48 \times 10^{19}$& $ 3.36 \times 10^{19}$&  $ 3.36 \times 10^{19}$\\
19 & $-1.18 \times 10^{23}$ & $-1.36 \times 10^{21}$& $-1.32 \times 10^{21}$ &  $-1.36 \times 10^{21} $&  $-1.36 \times 10^{21}$\\
20 & $-5.01 \times 10^{24}$ &  $5.90 \times 10^{22}$ &  $6.07 \times 10^{22}$& $5.92 \times 10^{22}$& $5.92 \times 10^{22}$\\
21 & $-2.34 \times 10^{26}$ &  $-2.68 \times 10^{24}$& $-2.62 \times 10^{24}$ & $-2.68 \times 10^{24}$ & $-2.68 \times 10^{24}$\\ 
 22 & $-1.09 \times 10^{28}$& $1.28 \times 10^{26}$& $1.31 \times 10^{26}$& $1.28 \times 10^{26}$& $1.28 \times 10^{26}$\\
23& $-5.54 \times 10^{29}$&$-6.41 \times 10^{27}$&  $-6.32 \times 10^{27}$& $-6.41 \times 10^{27}$&  $-6.41 \times 10^{27}$\\
24 &$-2.80 \times 10^{31}$& $3.35 \times 10^{29}$& $3.39 \times 10^{29}$& $3.35 \times 10^{29}$&  $3.35 \times 10^{29}$\\
25 & $-1.54 \times 10^{33}$& $-1.83 \times 10^{31}$& $-1.81 \times 10^{31}$& $-1.83 \times 10^{31}$& $-1.83 \times 10^{31}$\\[0.04cm]
\hline\hline \end{tabular}
\end{table*}

\section{ Higher-order coefficients from the Adler function in $\overline{\text{MS}}$ scheme}\label{sec:DMSbar}
We recall that the aim of this work is to make predictions on the higher-order perturbative coefficients $c_{n,1}$ with $n\ge 5$, using as input the known coefficients $c_{n,1}$ with $n\leq 4$ given in (\ref{eq:cn1}). As these coefficients appear in the expansion (\ref{eq:B}) of the Borel transform,  we focus on this function. We start from the remark that this function  admits a Taylor series convergent in a disk around the origin of the complex Borel plane and  propose alternative expansions, which converge in a larger domain and implement in an optimal way the known  analyticity properties in the  Borel plane. When reexpanded in powers of the Borel variable $u$, these expansions contain higher-order terms, which allow the extraction of the perturbative coefficients of interest. 

Specifically, the strategy involves the following algorithmic steps, which we explain in detail using for illustration the optimal expansion (\ref{eq:Bw1}): assuming we know $N$ coefficients $c_{n,1}$ for $1\leq n\leq N$,  we start from the  expansion (\ref{eq:B}) of $B_D(u)$ truncated at a finite order $N-1$. We insert $u=\tilde u(w)$  in this truncated expansion and expand its product with the global prefactor $(1+w)^{2\gamma_1} (1-w)^{2\gamma_2}$ in powers of $w$  to the same order $N-1$. This gives a polynomial in $w$ of order $N-1$, with $N$ known nonzero coefficients $f_n$ for  $0\leq n\leq N-1$. Finally, we reexpand in powers of $u$ the expression (\ref{eq:Bw1}), where the series in powers of $w$ is truncated at $n\leq N-1$. In this way, we recover the first $N$ input coefficients $c_{n,1}$ entering the coefficients $b_n$ by (\ref{eq:bn}),  but obtain also definite values for the higher-order coefficients  $c_{n,1}$ for $n> N$. The same steps are applied when considering the expansions (\ref{eq:Bw}), (\ref{eq:Bv}) and (\ref{eq:Bv1}).

\begin{table*}[htb]
\caption{The same as in Table \ref{tab:1} for the alternative model proposed in \cite{CaFi2011}, summarized in the Appendix.  } \vspace{0.1cm}
\label{tab:2}
 \renewcommand{\tabcolsep}{0.55pc} 
\renewcommand{\arraystretch}{1.15} 
\begin{tabular}{l|l l  |l l  | l}\hline\hline
$N$ \,\,\,    & Eq. (\ref{eq:Bv}) &Eq. (\ref{eq:Bw}) & Eq. (\ref{eq:Bv1})  & Eq. (\ref{eq:Bw1})& Exact $c_{N,1}$    \\\hline
4 & $-52.34$ &$ -17.61$ & 14.77 & 17.85 & 49.076 \\
5 & $-932.45 $& $-270.46$ & 255.98& 255.73 & 283. \\
6 & $-14348.46$ &$-2290.94$ & 3096.35 & 2928.76 &  2654.51\\
7 & $-253427.4$ &$-28576.4$ & 21587.4&  18171.5&  7901.76 \\
8 &$ -4.16\times 10^6$ &14826.9 & 470224.4 & 322587.1 & 241607.9\\
9 &$ - 8.18\times 10^7$ &$-2.03\times 10^6$ & $1.77\times 10^6 $ &$ -1.48 \times 10^6$&  $-982236.7$\\
10 & $-1.57 \times 10^9$& $4.93 \times 10^7$ & $8.13 \times 10^7$ & $ 4.21 \times 10^7$&   $5.85 \times 10^7$\\
11 &$ -3.67 \times 10^{10}$ & $-1.10 \times 10^9$ & $-8.97 \times 10^8$ & $-9.95 \times 10^8$ & $-8.69 \times 10^8$\\
12 &$-8.36 \times 10^{11}$& $2.31 \times 10^{10}$ &  $2.09\times 10^{10}$& $2.95 \times 10^{10}$ &  $2.86 \times 10^{10}$\\
13 & $-2.35 \times 10^{13}$ & $-7.89 \times 10^{11}$ & $-9.58 \times 10^{11}$ & $-6.52 \times 10^{11}$ & $-6.84 \times 10^{11}$\\
14 &$-6.32 \times 10^{14}$ & $2.05 \times 10^{13}$ &  $1.43 \times 10^{13}$ & $2.24 \times 10^{13}$& $ 2.21 \times 10^{13}$\\
15 &$ -2.10 \times 10^{16}$ & $-6.98 \times 10^{14}$ & $-8.81 \times 10^{14}$ & $-6.77 \times 10^{14}$& $-6.76 \times 10^{14}$\\
16 &$ -6.58 \times 10^{17}$ & $2.35 \times 10^{16}$&  $1.84 \times 10^{16}$& $2.37 \times 10^{16}$ &  $2.37 \times 10^{16}$\\
17 &$ -2.53 \times 10^{19}$& $-8.59 \times 10^{17}$& $-9.92 \times 10^{17}$&  $-8.56 \times 10^{17}$& $-8.55 \times 10^{17}$\\
18 & $-9.03 \times 10^{20}$ & $3.33 \times 10^{19} $&  $ 2.99 \times 10^{19}$& $ 3.34 \times 10^{19}$&  $ 3.34 \times 10^{19}$\\
19 & $-3.94 \times 10^{22}$ & $-1.36 \times 10^{21}$& $-1.45 \times 10^{21}$ &  $-1.36 \times 10^{21} $&  $-1.36 \times 10^{21}$\\
20 &$ -1.57 \times 10^{24}$ &  $5.90 \times 10^{22}$ &  $5.71 \times 10^{22}$& $5.91 \times 10^{22}$& $5.91 \times 10^{22}$\\
21 & $-7.64 \times 10^{25}$ &  $-2.69 \times 10^{24}$& $-2.72 \times 10^{24}$ & $-2.68 \times 10^{24}$ & $-2.68 \times 10^{24}$\\ 
 22 &$ -3.36 \times 10^{27}$& $1.28 \times 10^{26}$& $1.28 \times 10^{26}$& $1.28 \times 10^{26}$& $1.28 \times 10^{26}$\\
23& $-1.79 \times 10^{29}$& $-6.41 \times 10^{27}$&  $-6.38 \times 10^{27}$& $-6.41 \times 10^{27}$&  $-6.41 \times 10^{27}$\\
24 &$-8.59 \times 10^{30}$& $3.35 \times 10^{29}$& $3.38 \times 10^{29}$& $3.35 \times 10^{29}$&  $3.35 \times 10^{29}$\\
25 &$ -4.99 \times 10^{32}$& $-1.83 \times 10^{31}$& $-1.81 \times 10^{31}$& $-1.83 \times 10^{31}$& $-1.83 \times 10^{31}$\\[0.04cm]
\hline\hline \end{tabular}
\end{table*}

Before presenting our results for the first unknown perturbative coefficients, it is instructive to investigate the potential of the four expansions  (\ref{eq:Bw}), (\ref{eq:Bw1}), (\ref{eq:Bv}) and (\ref{eq:Bv1}) to predict the next perturbative coefficient $c_{N,1}$ from the knowledge of the coefficients $c_{n,1}$ with $n\leq N-1$, for increasing orders $N$. The exercise is motivated by the remark that,  if a function  is expanded as a convergent power series,   the knowledge of an increasing number of expansion coefficients is expected to strongly constrain  the next terms, which should be close to the exact terms of the full expansion. 

 For generating higher-order coefficients we used first a model of the Adler function proposed in \cite{BeJa}, which we summarize for completeness in the Appendix.
In Table \ref{tab:1}, we present the results given by the expansions  (\ref{eq:Bw}), (\ref{eq:Bw1}), (\ref{eq:Bv}) and (\ref{eq:Bv1}), compared to the exact coefficients of the model, given in the last column. 

We first note that  for the best expansion  (\ref{eq:Bw1}) in powers of the optimal mapping with the exact implementation of the nature of the first singularities,  the predicted coefficients $c_{N,1}$ listed in column 5 start to be close to the exact values for  $N\ge 5$ and practically coincide with them  for higher $N$ (the number of identical digits in the corresponding values is actually larger than is shown by rounding). By comparison, as shown in column 3, the expansion (\ref{eq:Bw}) in powers of the optimal mapping without the implementation of the nature of the nearest singularities has a poor predictive power of the next coefficient at low orders, but gradually approaches the exact results at high orders, confirming the asymptotic convergence rate of this expansion,  mentioned above.  

In Table \ref{tab:1} we show also the predictions of the expansions (\ref{eq:Bv}) and (\ref{eq:Bv1}) in powers of the mapping (\ref{eq:v}) proposed in  \cite{Mueller1985}. From the results given in column 2, one can see that the series  (\ref{eq:Bv}) fails to reproduce the next terms of the expansion from the knowledge of the previous ones. This is explained by the fact that the IR cut (the segment on the real axis shown in the right panel of Fig. \ref{fig:confBorel}) restricts the convergence of the series to a rather small domain. By softening the first singularities, as done in   (\ref{eq:Bv1}), the limitation set by the IR cut on the convergence is reduced and the predictive power of the expansion increases. As seen from column 4 of Table \ref{tab:1}, at high orders which are influenced by the first UV renormalon, the expansion reproduces with high accuracy the exact values of the next coefficients.

As remarked in the literature \cite{Pich2010, DeMa, CaFi2011, Abbas:2013usa}, the model  proposed in \cite{BeJa}  is characterized by a relatively large value of the first IR renormalon residue $d_2^{\rm IR}$.  It is of interest to consider also   alternative models with a smaller  residue. An example, proposed in \cite{CaFi2011}, is briefly presented in the Appendix. As seen from the perturbative coefficients listed in Eq. (\ref{eq:altcn1model}), the oscillatory character of the series, imposed by the UV renormalons, starts a bit earlier in this case compared to the previous model. On the other hand, the large-order behavior of the two models is the same, being dictated by the first UV renormalon which is modeled in the same way.

In Table \ref{tab:2} we present the same analysis as in Table \ref{tab:1}, performed for the alternative model.  The results given in columns 3 and 5 show that  the expansions (\ref{eq:Bw}) and (\ref{eq:Bw1}) based on the optimal conformal mapping (\ref{eq:w}) reproduce well the exact coefficient $c_{N,1}$ of the model at high orders. For the best expansion (\ref{eq:Bw1}), which softens the first singularities, the exact coefficients are reproduced also at low orders, although the approximation is slightly worse than for the previous model shown in Table \ref{tab:1}. For the nonoptimal mapping (\ref{eq:v}), the simple expansion  (\ref{eq:Bv}) fails to recover the next coefficient, while the expansion (\ref{eq:Bv1}) which softens the first singularities gives good results both  at large and intermediate orders.

Based on the above study, we shall choose the expansions (\ref{eq:Bw1}) and (\ref{eq:Bv1}) for predicting the higher coefficients from the known $c_{n,1}$ with $n\leq 4$ given in Eq. (\ref{eq:cn1}). One may argue that, as seen from the first rows of Tables \ref{tab:1} and \ref{tab:2}, these expansions are not able to recover the coefficient $c_{4,1}$ from the first three coefficients. However, since we do not use  {\em ad hoc} parametrizations, but systematic expansions with improved properties when the order is increased, we can expect the prediction of the coefficient $c_{5,1}$ and the next ones to be better. Moreover, in Sec. \ref{sec:Ioptim} we shall corroborate the predictions based on the expansion of the Adler function with those based on the expansion of a suitable weighted integral of $\wh D(s)$ in the complex $s$ plane. 

 Using the first four coefficients $c_{n,1}$ given in (\ref{eq:cn1}) and  applying the strategy explained at the beginning of this section, we arrive at the representation 
\beq\label{eq:Bwratio}
B_D(u)=\frac{1- 0.7973\, w + 0.4095 \,w^2 + 8.6647\, w^3}{(1+w)^{2\gamma_{1}} (1-w)^{2\gamma_{2}}},
\eeq
which, reexpanded in powers of $u$, reads
\bea\label{eq:Bwuseries}
B_D(u)&\!\!=\!\!&1 + 0.7288\, u + 0.6292\, u^2 + 0.7181\, u^3 \nn\\
&\!\!+\!\!& 0.4157\, u^4 +  0.4220\, u^5 + 0.1429\, u^6+\ldots
\eea
Using (\ref{eq:bn}), we recover from the first four coefficients the input values $c_{n,1}$ for $n\leq 4$, and from the remaining coefficients we predict
\beq\label{eq:c567w}
c_{5,1}=255.73, \,\,\,\, c_{6,1}=2920.2,\,\,\,\, c_{7,1}=13357.1\,.
\eeq
We note that the value of $c_{5,1}$ was already reported in Ref. \cite{CaFi2009}, where the representation (\ref{eq:Bw1}) was used for the extraction of the strong coupling from the hadronic $\tau$ width.

For the nonoptimal mapping (\ref{eq:v}), the representation analogous to (\ref{eq:Bwratio}) has the form
\beq\label{eq:Bvratio}
B_D(u)=\frac{1- 4.2947\, v + 1.6923\, v^2 + 32.1202\, v^3 }{(1+v)^{2\gamma_{1}} (1-v/\tilde (2))^{\gamma_{2}}},
\eeq
which, reexpanded in powers of $u$, leads to 
\bea\label{eq:Bvuseries}
B_D(u)&\!\!=\!\!&1 + 0.7288\, u + 0.6292\, u^2 + 0.7181\, u^3 \nn\\
&\!\!+\!\!& 0.4162\, u^4 + 0.4561 \, u^5 + 0.1397\, u^6+\ldots
\eea
The first four Taylor coefficients of this series coincide with those of the expansion (\ref{eq:Bwuseries}), being fixed by the values (\ref{eq:cn1}) used as input, while from the remaining ones we obtain
\beq\label{eq:c567v}
c_{5,1}=255.98, \,\,\, \, c_{6,1}=3156.4,\,\,\,\, c_{7,1}=13047.8\,.
\eeq
\section{ Higher-order coefficients from the Adler function in $C$ scheme}\label{sec:DC}
As remarked in Sec. \ref{sec:Adler},  the nature of the singularities of the Borel transform in the $u$ plane depend only on the first two coefficients, $\beta_1$ and $\beta_2$, of the $\beta$ function, which are  scheme independent. This means that the first singularities of the function $\widehat B(u, C)$  defined in (\ref{eq:BhatC}) are expected to have the same location at $u=-1$ and $u=2$, and their nature to be described by the same relations (\ref{eq:gammapowers}). Another argument in favor of this property, put forward in \cite{Mueller1985}, is that the behavior of the Borel transform near the first singularities is dictated by the limit of  vanishing coupling, where the $\overline{\text{MS}}$ and the $C$ scheme coincide. Therefore,  we can adopt for the function $\widehat B(u, C)$ defined in (\ref{eq:BhatC}) the expansions written in (\ref{eq:Bw1}) and (\ref{eq:Bv1}). 

In particular, we consider the expansion
\be\label{eq:Bw1C}
 \widehat B(u, C)=\frac{1}{(1+w)^{2\gamma_{1}} (1-w)^{2\gamma_{2}}} \sum_{n\ge 0} \hat f_n(C)\, w^n,
\ee
 based on the optimal mapping (\ref{eq:w}) and the softening of the first singularities. The expansion (\ref{eq:Bw1C})  is similar to (\ref{eq:Bw1}), the only difference being that now the coefficients $\hat f_n$  depend on $C$.

Using as input the first four coefficients $\hat c_{n, 1}(C)$ from (\ref{eq:hatDC1}) and applying the steps presented in the previous section, we arrive at the representation
\bea\label{eq:BCwratio}
&&\hspace{-0.4cm} \widehat B(u, C)=\frac{1} {(1+w)^{2\gamma_{1}} (1-w)^{2\gamma_{2}}} \left[1 - (0.797 - 2.667\, C) w \right.\nn\\
&& + \,(1.333 + 2.461\, C +   3.556\, C^2) w^2 \\
&& \left. + \, (10.69 + 2.306\, C + 8.149\,C^2 + 
    3.16\, C^3)\, w^3\right].\nn
\eea
Reexpanded in powers of $u$, this gives
\bea\label{eq:BCu}
&&\hspace{-0.1cm}\widehat B(u, C)=1 + (0.729 +  C) u + (0.759 + 1.124\, C \nn\\ 
&&\hspace{-0.4cm} +\, 0.5\, C^2) u^2+ (0.893 + 1.055\, C + 0.694\, C^2 + 
    0.167\, C^3) u^3 \nn\\
&& \hspace{-0.4cm} +\, (0.544 + 0.638\, C + 0.499\, C^2 +   0.046\, C^3) u^4 +\ldots
\eea

\begin{figure}[htb]
\includegraphics[scale=0.25]{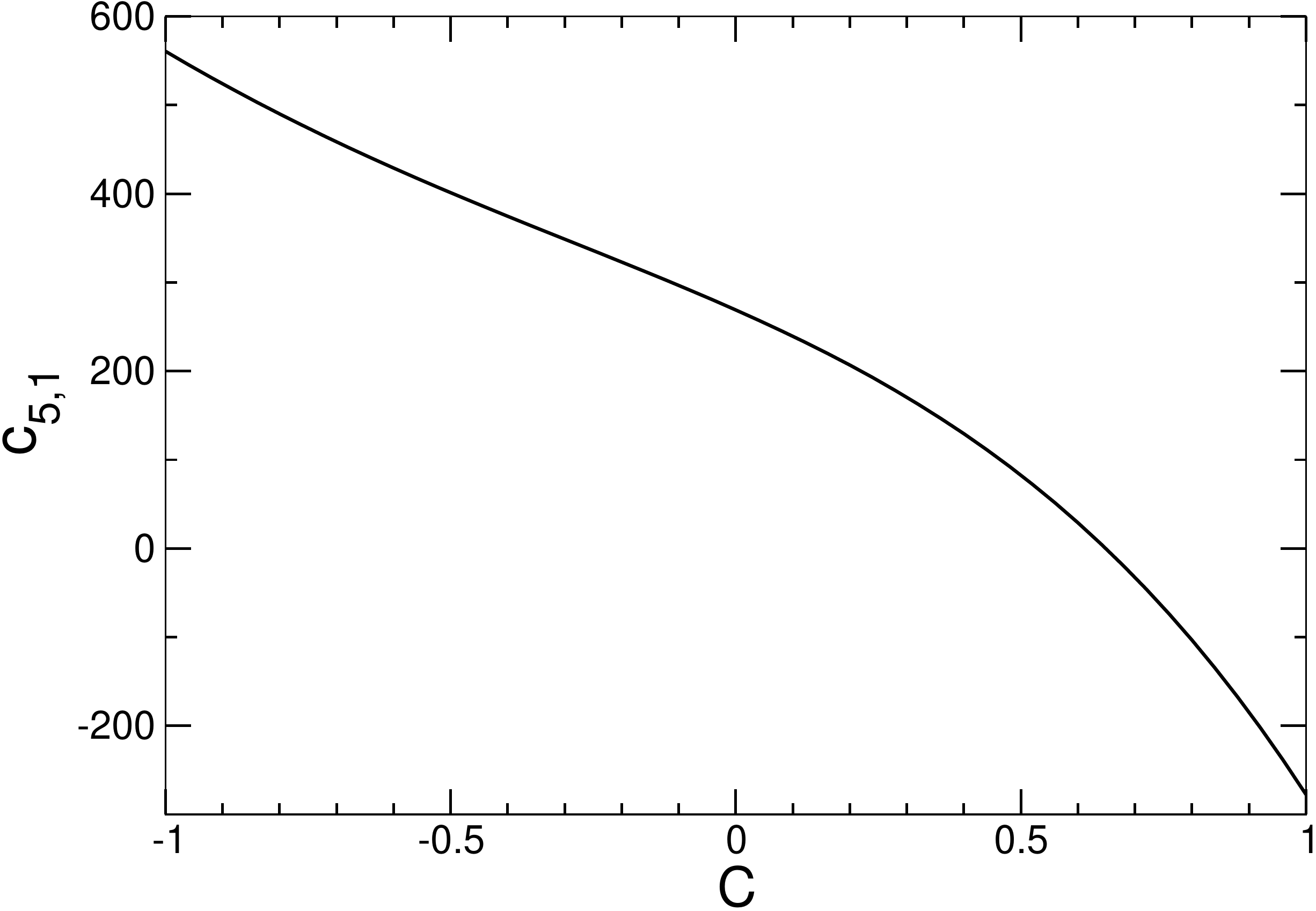}
\caption{ The coefficient $c_{5,1}$ given by (\ref{eq:c51C}) plotted as a function of $C$. \label{fig:2}}
\end{figure}

Using Eq. (\ref{eq:hatbn}), one can check that the first four terms of the expansion (\ref{eq:BCu}) reproduce exactly the known coefficients of the expansion (\ref{eq:hatDC1}) of the Adler function, while from the last term  of  (\ref{eq:hatbn}) compared with the last term of (\ref{eq:hatDC1}) we extract the unknown perturbative coefficient $c_{5,1}$ in the $\overline{\text{MS}}$ scheme as
\beq\label{eq:c51C}
c_{5,1}=268.9 - 285.1\, C - 101.8\, C^2 - 133.9\, C^3 - 25.6\, C^4.
\eeq

  We plot this expression in Fig. \ref{fig:2}, for the parameter $C$ in the interval from $-1$ to 1. A very similar curve is obtained using an expansion  based on the alternative mapping (\ref{eq:v}). 

 Figure \ref{fig:2} shows a quite drastic variation with $C$ of the perturbative coefficient $c_{5,1}$, which actually must be independent of $C$.  As discussed in the previous studies \cite{Cscheme, Caprini:2018agy}, a reasonable range of $C$ appears to be situated close to the origin. By restricting for instance $C$  to the range 
from $-0.05$ to $0.05$, we obtain $c_{5,1}\in (269,\,401)$, with  $c_{5,1}=283$ for $C=0$, values which are consistent with the determinations  in Sec. \ref{sec:DMSbar}.  However, since no prescription for choosing a narrow range of $C$  exists and Fig. \ref{fig:2}  does not indicate a region of stability,  we conclude that the $C$ scheme is not useful for an accurate extraction of the higher-order perturbative coefficients of the Adler function.  

\section{Higher-order coefficients from $\tau$ hadronic width}\label{sec:delta0}
 In order to apply the method of conformal mappings to the expansion of $\tau$ hadronic width,  the  analytic properties of the  Borel transform $B_\delta(u)$ defined in (\ref{eq:Bdelta}) must be known. Information on these properties can be obtained by establishing a relation between the functions $B_\delta$ and $B_D$. This relation was investigated in several works (see for instance \cite{Brown:1992pk, BeJa, Boito:2018rwt}).

 If we introduce  the Laplace-Borel representation (\ref{eq:Laplace}) in the integral (\ref{eq:delta0}) and permute the integrals we obtain
\beq\label{eq:integralphi}
\delta^{(0)} = \frac{1}{\beta_0} \,\int\limits_0^\infty  d u\,  \,  B_D(u)\, \frac{1}{2\pi}\int\limits_{0}^{2\pi} \, d\phi\,
e^{\frac{-u}{\beta_0 a(-s)}}  (1-e^{i\phi})^3\,(1+ e^{i\phi}),
\eeq
where $-s=m_\tau^2 \exp(i(\phi-\pi))$.

 The integral upon $\phi$ can be performed exactly in the one-loop approximation of the running coupling, when (\ref{eq:rge}) implies
\beq\label{eq:oneloop}
\frac{1}{\beta_0 a(-s)}=\frac{1}{\beta_0 a(m_\tau^2)} + \ln\frac{-s}{m_\tau^2}, 
\eeq
where the last term is equal to $i(\phi-\pi)$. Then, the comparison with (\ref{eq:Laplacedelta0}) leads to \cite{BeJa}
\beq\label{eq:BdeltaBD}
B_\delta(u)=\frac{12}{(1-u)(3-u)(4-u)} \frac{\sin(\pi u)}{\pi u} B_D(u).
\eeq
 
From (\ref{eq:BdeltaBD}) it follows that $B_\delta(u)$ inherits from  $B_D(u)$ the first singularities at $u=-1$ and $u=2$. No new singularities appear, the  poles at $u=0$, 1, 3 and 4 being canceled by the zeros of $\sin(\pi u)$. So, we can apply to $B_\delta(u)$ the method of conformal mappings, using the same optimal variable $w$ defined in (\ref{eq:w}). We notice further that  $\sin(\pi u)$  exhibits also simple zeros at $u=-1$ and $u=2$, which reduce by 1 the strength of  the singularities of   $B_D(u)$ given in (\ref{eq:gammapowers}). So, we can use for $B_\delta(u)$ an expansion similar to
(\ref{eq:Bw1}), with exponents in the prefactors smaller by 1. Using the first four known coefficients (\ref{eq:dn}) and the strategy presented in Sec. \ref{sec:DMSbar}, we obtain the representation
\beq\label{eq:Bdeltaw1}
B_\delta(u)=\frac{1 + 3.425\, w + 7.695\, w^2 + 7.189\, w^3}{(1+w)^{2(\gamma_{1}-1)} (1-w)^{2(\gamma_{2}-1)}},
\eeq
which, reexpanded in powers of $u$, leads to
\bea
B_\delta(u)&=&1 + 2.312 u + 2.604 u^2 + 1.859 u^3\nn\\
& +& 1.114 u^4 + 
 0.694 u^5 + 0.360 u^6+\ldots
\eea
The first four terms reproduce the coefficients $b_n'$ of the expansion (\ref{eq:Bdelta}), known from (\ref{eq:dn}) and (\ref{eq:bnprime}),  while from the next three terms we extract the  coefficients
\beq\label{eq:c567deltaw}
c_{5,1}=378, \,\,\,\, c_{6,1}= 3922,\,\,\,\, c_{7,1}=24414\,.
\eeq

We recall that the expression (\ref{eq:BdeltaBD}) is only approximate: beyond one loop, one expects the simple zeros of $\sin(\pi u)$ to be replaced by branch points which vanish at the relevant points and modify the prefactors in the representation (\ref{eq:Bw1}) of $B_D(u)$ by a certain unknown amount. The values (\ref{eq:c567deltaw}) obtained in the limit of one-loop coupling can be viewed therefore only as a qualitative prediction.

To obtain further insight into the problem, we include in the renormalization-group equation (\ref{eq:rge})  the two-loop term in the $\beta$ function. Then (\ref{eq:oneloop}) is modified to \cite{Brown:1992pk}
 \beq\label{eq:twoloop}
\frac{1}{\beta_0 a(-s)}=\frac{1}{\beta_0 a(m_\tau^2)} + \ln\frac{-s}{m_\tau^2}-\frac{\beta_2}{2\beta_0^2}\ln\frac{a(-s)}{a(m_\tau^2)}.
\eeq
In an iterative approach, we use again (\ref{eq:oneloop}) in order to evaluate the last term, which we then expand to order $a(m_\tau^2)$, to obtain: 
\beq\label{eq:twoloop1}
\frac{1}{\beta_0 a(-s)}=\frac{1}{\beta_0 a(m_\tau^2)} + i(\phi-\pi) \left[1+\frac{\beta_2}{2\beta_0}\frac{\alpha_s(m_\tau^2)}{\pi}\right].
\eeq
The integration upon $\phi$ in (\ref{eq:integralphi}) can be done exactly also in this case by a simple rescaling of the variable $u$, and we obtain instead of (\ref{eq:BdeltaBD}) the relation 
\beq\label{eq:BdeltaBD1}
B_\delta(u)=\frac{12}{(1-u \xi)(3-u\xi)(4-u \xi)} \frac{\sin(\pi u \xi)}{\pi u \xi} B_D(u),
\eeq
where
\beq\label{eq:xi}
\xi=1+\frac{\beta_2}{2\beta_0}\frac{\alpha_s(m_\tau^2)}{\pi}= 1+ 0.5659\, \alpha_s(m_\tau^2).
\eeq

Of course, the relation (\ref{eq:BdeltaBD1}) is only an approximation, since $B_\delta(u)$ has some residual dependence on the
coupling  $\alpha_s(m_\tau^2)$ contained in the parameter $\xi$. However, the expression (\ref{eq:xi}) shows that for current values $\alpha_s(m_\tau^2)\sim 0.3$ the parameter $\xi$ differs from 1 by a small quantity. Therefore, its presence in the sine function leads to only a slight shift of the position of the zeros. In particular, instead of zeros at $u=2$ and $u=-1$, this factor vanishes at the nearby points $u=2/\xi$ and $u=-1/\xi$. In consequence, the strength of the singularities of $B_D(u)$ at $u=2$ and $u=-1$ is not modified in a manifest way, but is indirectly attenuated by the simple zeros that $B_\delta(u)$ is expected to have near these points. 

Using the above discussion,  we consider a representation of $B_\delta(u)$ of the form
\beq\label{eq:Bdeltaw2}
B_\delta(u)=\frac{(w-\tilde w(-1/\xi)) (w-\tilde w(2/\xi))}{(1+w)^{2\gamma_{1}} (1-w)^{2\gamma_{2}}}\,\sum_{j=0}^3 g_j w^j,
\eeq
which exhibits simple zeros as the positions indicated above. The coefficients $g_j$ are fixed by the condition of reproducing the first four known Taylor coefficients in the standard expansion (\ref{eq:Bdelta}), and from the higher-order  terms we predict the higher-order perturbative coefficients.

\begin{figure}[htb]
\includegraphics[scale=0.25]{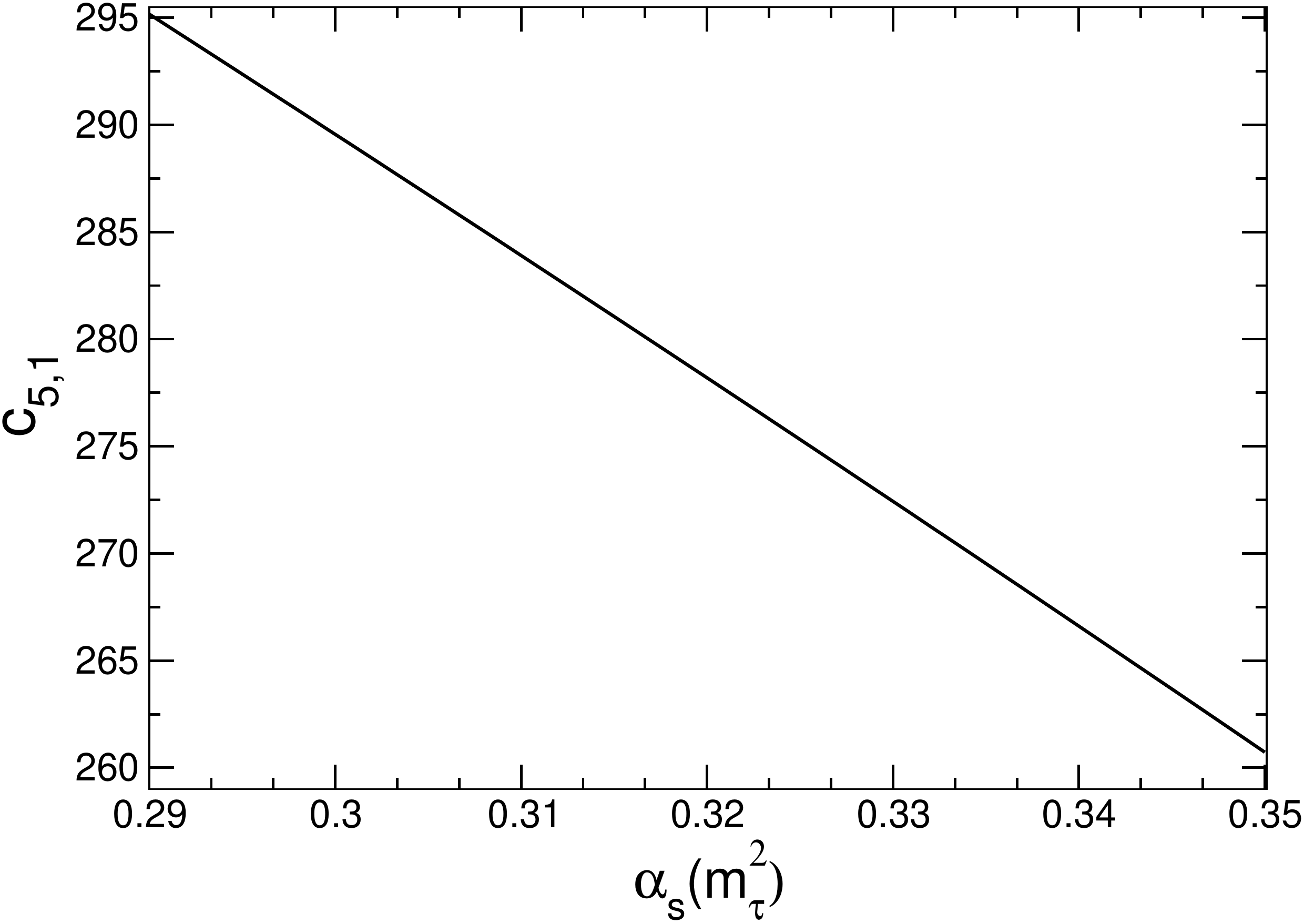}
\caption{ The coefficient $c_{5,1}$ obtained from the representation (\ref{eq:Bdeltaw2}), for various values of $\alpha_s(m_\tau^2)$ in the parameter $\xi$. \label{fig:3}}
\end{figure}

For illustration, we present in Fig. \ref{fig:3} the coefficient $c_{5,1}$ calculated from  the representation (\ref{eq:Bdeltaw2}) for a physical range of values  of $\alpha_s(m_\tau^2)$. The comparison with (\ref{eq:c567deltaw}) shows that the inclusion of higher-loop effects in the running coupling shifts the predicted value of $c_{5,1}$ towards smaller values, consistent with the predictions made in Sec. \ref{sec:DMSbar}. However, since the frame in which we worked is only approximate, we consider this prediction  only as a qualitative insight towards the exact result.

\section{Other contour integrals}\label{sec:weights} 
The analysis  presented in Sec. \ref{sec:DMSbar} proved that the power of the method of conformal mapping is increased if  the nature of the first singularities of the Borel transform is known.  For the $\tau$ hadronic width this information is not exactly available. In the one-loop (large-$\beta_0$) limit, the factor connecting $B_\delta(u)$ to $B_D(u)$ contains simple zeros at $u=2$ and $u=-1$, which modify the nature of the first singularities.  A hint about what happens if we go beyond one loop was provided in the previous section. But the nature of the singularities in the exact case remains unknown.  Therefore, the $\tau$ hadronic width is not a suitable observable for predicting 
   the higher-order perturbative coefficients  with the method of conformal mappings applied in this paper.

It is of interest to look for other quantities for which the first singularities of the Borel transform can be exactly determined. One may think to consider, instead of (\ref{eq:delta0}), more general contour integrals of the form
\be\label{eq:Iomega}
I_\omega =  \frac{1}{2\pi i} \oint\limits_{|s|=m_\tau^2}\, \frac{d s}{s} \,\omega(s)\, \wh D(s),
\ee
where $\omega(s)$ is a suitable weight. 

 A large class of integrals of this form have been investigated in \cite{BBJ, Abbas:2013usa} for testing the perturbative expansions of the moments of the spectral function ${\rm Im}\Pi(s)$. In these analyses, the weights [denoted as $W_i(s)$ in \cite{Abbas:2013usa}], must be boundary values of analytic functions in the disk $|s|<m_\tau^2$, in order to connect  by Cauchy relation the contour integrals to observables measured on the timelike axis.  In the present frame, this restriction is not necessary, since we only look for the perturbative expansion of the quantity $I_\omega$. We retain however the requirement that $\omega(s)$ vanish at the timelike point $s=m_\tau^2$, in order to suppress the contribution of the region where the perturbative  logarithms in (\ref{eq:hatD}) are large, worsening the convergence of the expansion.

 We are actually interested in weights $\omega(s)$ for which  the singularities of the corresponding Borel transform $B_{I_\omega}(u)$ in the $u$ plane can be inferred with some confidence. We investigate the problem by using insight from the limit of one-loop coupling. In this limit,  $B_{I_\omega}(u)$ is given by a relation of the form 
\beq\label{eq:F}
B_{I_\omega}(u)=F_\omega(u) B_D(u),
\eeq
where $F_\omega(u)$ is a calculable function containing explicitly the factor $\sin(\pi u)$ as in (\ref{eq:BdeltaBD}).  

Several restrictions must be imposed on this function, in order to ensure suitable properties for  $B_{I_\omega}(u)$. First, we require that $F_\omega(u)$ do not vanish at $u=2$ and $u=-1$. The reason of this condition is simple:   in the limit of one-loop coupling, as will be seen below, $F_\omega(u)$ has only simple zeros.  But in the exact case,  a simple zero is expected to become a branch point. Therefore, the zeros at $u=-1$ and $u=2$ would change the nature of the singularities present in $B_D(u)$ at these points by an unknown amount, introducing an uncertainty in the behavior of  $B_{I_\omega}(u)$. On the other hand,  no branch points  at  $u=2$ and $u=-1$ are expected to appear if zeros in  $F_\omega(u)$  at these points are absent in the one-loop  limit. So, if $F_\omega(u)$ does not vanish at $u=2$ and $u=-1$, we can say with some confidence that the singularities of  $B_{I_\omega}(u)$ at these points have the same nature as those of $B_D(u)$. 
We require also that  $F_\omega(u)$ do not exhibit zeros at low values of $|u|$, in particular on the interval $(-1,\, 2)$, since in the exact case the simple zeros due to the sine function are expected to become branch points, which modify the analytic properties of $B_{I_\omega}(u)$.

  We investigated a large class of weights $\omega(s)$, for which we calculated explicitly the function $F_\omega(u)$. 
 For illustration, we present several choices in Table \ref{tab:3}, where at $i=5$ we give for completeness the weight corresponding to the physical quantity $\delta^{(0)}$. 

It turns out that the above conditions restrict seriously the choice of acceptable weights. One can see that for $i\leq 6$, the functions $F_{\omega_i}(u) $ vanish at $u=-1$, since there is no factor in the denominator to compensate the zero of $\sin (\pi u)$. 
On the other hand, for $i=1,\,2,\,5,\,7$ and 8, the functions $F_{\omega_i}(u) $ vanish at $u=2$. For $i=10$, where the zeros of $\sin (\pi u)$ at $u=-1$ and $u=2$ are compensated by the denominator, there is still a zero at $u=1$, which is not acceptable since  in the exact case it will become an unwanted branch point below $u=2$. 

We conclude that, from the functions listed in Table 
\ref{tab:3}, only $\omega_9(s)$ satisfies the requirements discussed above. The corresponding $B_{I_\omega}(u)$ is expected to have the same analyticity properties in the $u$ plane as $B_D(u)$.  Therefore, this weight appears to be a suitable choice for the determination of the higher-order  coefficients from the perturbative expansion of $B_{I_\omega}(u)$. This determination will be presented in the next section.

\begin{table}[htb]
\caption{The function $F_\omega(u)$ defined in (\ref{eq:F}) for several weights $\omega_i(s)$. } \vspace{0.1cm}
\label{tab:3}
 \renewcommand{\tabcolsep}{0.55pc}  
\renewcommand{\arraystretch}{1.4} 
\begin{tabular}{c|l|l }\hline\hline
$i$&~~~~~~~~~$\omega_i(s)$ \,\,\,    &~~~~~~~~~ $F_{\omega_i}(u) $   \\\hline\hline
1&$ \left(1-\frac{s}{m_\tau^2}\right) $ &$\frac{1}{(1-u)} \frac{\sin(\pi u)}{\pi u}$\\[0.1cm]\hline
2&$ \left(1-\frac{s}{m_\tau^2}\right)^2 $ &$\frac{2}{(1-u)(2-u)} \frac{\sin(\pi u)}{\pi u}$\\[0.1cm]\hline
3&$ \left(1-\frac{s}{m_\tau^2}\right)^2\,\left(2+\frac{s}{m_\tau^2}\right) $ &$\frac{6}{(1-u)(3-u)} \frac{\sin(\pi u)}{\pi u}$\\[0.1cm]\hline
4&$ \left(1-\frac{s}{m_\tau^2}\right)^3 $ &$-\frac{6}{(1-u)(2-u)(3-u)} \frac{\sin(\pi u)}{\pi u}$\\[0.1cm]\hline
5&$ \left(1-\frac{s}{m_\tau^2}\right)^3\,\left(1+\frac{s}{m_\tau^2}\right) $ &$\frac{12}{(1-u)(3-u)(4-u)} \frac{\sin(\pi u)}{\pi u}$\\[0.1cm]\hline
6&$ \left(1-\frac{s}{m_\tau^2}\right)^3\,\left(3+\frac{s}{m_\tau^2}\right) $ &$\frac{24}{(1-u)(2-u)(4-u)} \frac{\sin(\pi u)}{\pi u}$\\[0.1cm]\hline
7&$ \left(1-\frac{s}{m_\tau^2}\right)\,\frac{m_\tau^2}{s} $ &$-\frac{1}{(1+u)} \frac{\sin(\pi u)}{\pi u}$\\[0.1cm]\hline
8&$ \left(1-\frac{s}{m_\tau^2}\right)^2\,\frac{m_\tau^2}{s} $ &$-\frac{2}{(1-u)(1+u)} \frac{\sin(\pi u)}{\pi u}$\\[0.1cm]\hline
9&$ \left(1-\frac{s}{m_\tau^2}\right)^3\,\frac{m_\tau^2}{s} $ &$-\frac{6}{(1-u)(2-u)(1+u)} \frac{\sin(\pi u)}{\pi u}$\\[0.1cm]\hline
10&$ \left(1-\frac{s}{m_\tau^2}\right)^3\,\left(1+\frac{s}{m_\tau^2}\right)\frac{m_\tau^2}{s} $ &$-\frac{12}{(2-u)(3-u)(1+u)} \frac{\sin(\pi u)}{\pi u}$\\[0.1cm]\hline
\hline
\end{tabular}
\end{table}

\section{Higher-order coefficients from a suitable contour integral}\label{sec:Ioptim}
We consider the integral
\be\label{eq:Ioptim}
I =  \frac{1}{2\pi i} \oint\limits_{|s|=m_\tau^2}\, \frac{d s}{3s} \, \left(\frac{s}{m_\tau^2}-1\right)^3\,\frac{m_\tau^2}{ s}\, \wh D(s),
\ee
where a normalization factor was introduced for convenience.

The perturbative expansion of the quantity $I$ reads
\beq\label{eq:Ipert}
I=\sum_{n\ge 1} I_n \,a_\mu^n, \quad\quad \mu=m_\tau,
\eeq
where the first coefficients are
\bea\label{eq:In}
&&\hspace{-0.5cm}I_1=1,\, I_2= 2.76,\, I_3= 8.06,\, I_4=-17.85 + c_{4,1},\nn\\&&\hspace{-0.5cm}I_5=-379.33 + 4.5\, c_{4,1} + c_{5,1},\\&&\hspace{-0.5cm} I_6=
 -2190.8 - 31.99\, c_{4,1} + 5.63\, c_{5,1} +  c_{6,1},\nn\\
&&\hspace{-0.5cm} I_7=-895.7  - 406.2\, c_{4,1} - 49.98 \,c_{5,1} + 6.75 \,c_{6,1} + c_{7,1}.\nn
\eea
We have used in the calculation the first three coefficients $c_{n,1}$ from (\ref{eq:cn1}), and left free $c_{4,1}$ and the next coefficients.

From the discussion in the previous section, we expect the Borel transform of $I$, defined in analogy with (\ref{eq:B}) and (\ref{eq:bn}) by the Taylor series
\beq\label{eq:BI}
B_I(u)=\sum_{n= 0}^\infty \,\frac{I_{n+1}}{\beta_0^n n!}\,u^n,
\eeq
to have analyticity properties in the $u$ plane similar to those of the Borel transform $B_D(u)$ of the Adler function.  In particular, because the corresponding function $F_\omega$ appearing in (\ref{eq:F}) does not have zeros at $u=-1$ and $u=2$, the nature of the first singularities of $B_I(u)$ is expected to be given by (\ref{eq:gammapowers}) and (\ref{eq:gamma12}).
Therefore, we can represent $B_I(u)$ by an expansion in powers of the optimal variable $w$, with the implementation of the nature of the first singularities, similar to the expansion   (\ref{eq:Bw1}) of the Adler function. 

As a first check, we kept three terms in the numerator of the representation, using as input the first three coefficients given in (\ref{eq:In}). When reexpanded in powers of $u$, this representation contains higher terms, from which, using (\ref{eq:In}), we  extracted the five-loop coefficient 
\beq\label{eq:c41BIw}
c_{4,1}=53.3.
\eeq

Using then as input the first four coefficients from (\ref{eq:In}), with the known value of $c_{4,1}$ from (\ref{eq:cn1}), we obtained the representation
\beq\label{eq:BIwratio}
B_I(u)=\frac{1-0.536\, w - 1.168\, w^2 - 1.181\, w^3}{(1+w)^{2\gamma_{1}} (1-w)^{2\gamma_{2}}},
\eeq
which, reexpanded in powers of $u$, reads
\bea\label{eq:BIwuseries}
B_I(u)&\!\!=\!\!&1 + 1.229\, u + 0.796\, u^2 + 0.457\, u^3 \nn\\
&\!\!+\!\!&0.274 \, u^4 + 0.133 \, u^5 + 0.091\, u^6+\ldots
\eea
Using (\ref{eq:In}) and (\ref{eq:BI}), we obtained from this expansion the next perturbative coefficients
\beq\label{eq:c567BIw}
c_{5,1}=327.0, \,\,\, \, c_{6,1}=2840.6,\,\,\,\, c_{7,1}=26475\,.
\eeq

If we use, instead of (\ref{eq:Bw1}), the  expansion (\ref{eq:Bv1}) based on the alternative conformal mapping (\ref{eq:v}), the results are
\beq\label{eq:c41BIv}
c_{4,1}=51.6,
\eeq
and, respectively,
\beq\label{eq:c567BIv}
c_{5,1}=308.9, \,\,\, \, c_{6,1}=2876.0,\,\,\,\, c_{7,1}=22829\,.
\eeq
\section{Average of the unbiased predictions}\label{sec:final}
In the previous sections, we investigated the prediction of the higher-order perturbative coefficients $c_{n,1}$ using the method of conformal mappings for the expansions of the Adler function and of its contour integrals. The investigation in Sec. \ref{sec:DC} showed that a precise prediction using the $C$ renormalization scheme is not  possible, since an allowed interval for the parameter $C$ is not {\em a priori} available. Furthermore, the analysis presented in Sec. \ref{sec:delta0} showed that in the case of the $\tau$ hadronic width, the behavior  of the  Borel transform near the first singularities, which plays an important role in the method applied in this paper,  is not known exactly.  

Therefore, we retain for calculating an average the predictions  obtained from the expansions of the Adler function in the $\overline{\text {MS}}$ scheme, investigated in Sec. \ref{sec:DMSbar},  and the  contour integral considered in Sec. \ref{sec:Ioptim}. In these cases,  the nature of the first singularities in the Borel plane is exactly known, which considerably improves the predictive power of the method of conformal mappings.
For these quantities, we used both expansions (\ref{eq:Bw1}) and (\ref{eq:Bv1}), based on the optimal mapping (\ref{eq:w}) and the alternative mapping (\ref{eq:v}).

Assuming first that only three perturbative coefficients from (\ref{eq:cn1}) are used as input, the method leads to a prediction for the coefficient $c_{4,1}$. From the values given by the above expansion for $N=4$ in Table \ref{tab:1} and the results quoted in Eqs. (\ref{eq:c41BIw}) and (\ref{eq:c41BIv}), we obtain the average  
\beq\label{eq:c41final}
c_{4,1}=34.4\pm 19.6,
\eeq
where we took as error the largest of the up and down values. We note that the error is rather large, which is actually to be expected at such a low order. The prediction is however compatible within errors with the true value $c_{4,1}=49.076$ given in (\ref{eq:cn1}).
 
Using as input the first four coefficients from (\ref{eq:cn1}), the method leads to the predictions for the next coefficients given in Eqs.  (\ref{eq:c567w}), (\ref{eq:c567v}), (\ref{eq:c567BIw}) and (\ref{eq:c567BIv}). Taking the average of these values  we obtain 
\bea\label{eq:c567final}
&&c_{5,1}=287 \pm 40,\quad \quad c_{6,1}=2948 \pm 208,\nn\\
&& c_{7,1}=(1.89\pm 0.75)\times 10^4,
\eea
where, as above, the error is the largest of the up and down values. As in \cite{Boito:2018rwt}, we cannot attach a statistical meaning to this error. Rather, it is chosen such as to cover the range of the values entering the average.

\section{Summary and conclusions}\label{sec:conc}
The state of the art in perturbative QCD is the calculation of some correlators to five-loop order. For the Adler function, the known perturbative coefficients are given in (\ref{eq:cn1}).  The knowledge of the higher-order coefficients is of much interest, in particular for increasing the accuracy of the determination of the strong coupling $\alpha_s$ from hadronic $\tau$ decays.  As the exact calculations to six-loop order are not foreseen in the near future due to computational difficulties, various approximate estimates have been proposed recently. Of course, some theoretical information about the expanded function must be available  if one wants to say something about its higher-order Taylor coefficients.

In the present paper we exploited the analytic properties in the Borel plane, which encode the high-order behavior of the perturbative expansion. We proved that  the method of accelerating the series convergence by conformal mappings provides a useful tool for the present purpose. 

Specifically, we used the representations 
 (\ref{eq:Bw1}) and (\ref{eq:Bv1}) of the Borel transform, based on the optimal conformal mapping (\ref{eq:w}) and the alternative mapping (\ref{eq:v}), which implement also the known behavior near the first singularities. The first four perturbative coefficients (\ref{eq:cn1}) were used as input for fixing the first terms of the expansion in powers of the conformal mappings in the representations (\ref{eq:Bw1}) and (\ref{eq:Bv1}).  When reexpanded in powers of $u$, these expressions reproduce the known coefficients, but  contain also higher powers, from which the next coefficients can be extracted. The good performance of these expansions to predict higher-order coefficients has been tested in Sec.  \ref{sec:DMSbar} up to high orders, using two renormalon-based models of the Adler function summarized in the Appendix.

 We based our predictions on the expansion of the Adler function in the $\overline{\text{MS}}$ scheme discussed in Sec. \ref{sec:DMSbar},  and on the expansion of a suitable contour integral investigated in Sec. \ref{sec:Ioptim}. In both these cases the behavior of the Borel transform near the first singularities is known, this information being very useful for increasing the accuracy of the predictions. 
  Our final results given in Eq. (\ref{eq:c567final}) are obtained from the average of the four values given in Eqs.  (\ref{eq:c567w}), (\ref{eq:c567v}), (\ref{eq:c567BIw}) and (\ref{eq:c567BIv}),  with a conservative definition of the error.

It is of interest to compare these predictions with previous determinations made in the literature.
 In Ref.~\cite{BCK02} the value $c_{5,1}=
145\pm 100$ was suggested, using only partial information about the five-loop coefficient available at that time. The value obtained from the principle of FAC in Ref.~\cite{BCK08} is $c_{5,1}=275$, while  in Ref.~\cite{BeJa} the estimate $c_{5,1}
=283 \pm 142$ was made by studying the expansion of the $\tau$ hadronic width. Finally, we quote the most recent values $c_{5,1}=277 \pm 51$, $c_{6,1}=3460 \pm 690$ and
$c_{7,1}=(2.02\pm 0.72)\times 10^4$, obtained in \cite{Boito:2018rwt} from Pad\'e approximants of the expansion of the $\tau$ hadronic width.

Our predictions (\ref{eq:c567final}) are compatible with the above quoted values, in particular with the recent predictions made in Ref.  \cite{Boito:2018rwt}. It must be emphasized that the values obtained in \cite{Boito:2018rwt} and in the present paper are obtained with completely different methods, which strengthen the confidence in these values. Our results support therefore the statement made in \cite{Boito:2018rwt} that it seems unlikely that the six-loop  coefficient would not be within the intervals given above.

\subsection*{Acknowledgments}
I thank D. Boito for useful discussions. This work was supported by the Romanian Ministry of Research and Innovation, Contract No. PN 18090101/2018.

\appendix
\section{Mathematical models}
In order to assess the quality of various perturbative  frameworks,  the exact pattern of 
the higher-order  coefficients of the Adler function must be known. Since this knowledge is not available, a suitable ansatz is usually adopted.  
The description of the function in terms of its dominant singularities in the 
Borel plane is a natural choice, 
consistent  with the general principles of analyticity. 
However, a considerable ambiguity still remains  
because, while  the position and nature of the leading 
singularities are known theoretically,  nothing can be said from theory  
about their strengths. In \cite{BBJ}, some arguments in favor of a ``reference model" proposed in \cite{BeJa}  were put forth. 
This model seems to be natural because the residues of the first renormalons result from the fit of the known low-order coefficients and are not imposed by hand.

 The model  \cite{BeJa}  expresses the Borel transform $B_D(u)$ in terms of a few UV and IR renormalons:
\be\label{eq:BBJ}
B_D(u)=B_1^{\rm UV}(u) +  B_2^{\rm IR}(u) + B_3^{\rm IR}(u) +d_0^{\rm PO} + d_1^{\rm PO} u,
\ee
 where
\be\label{eq:BIR}
B_p^{\rm IR}(u)= \frac{d_p^{\rm IR}}{(p-u)^{\gamma_p}}\,
\left[\, 1 + \tilde b_1 (p-u)  +\ldots \,\right],
\ee
\be\label{eq:BUV}
B_p^{\rm UV}(u)=\frac{d_p^{\rm UV}}{(p+u)^{\bar\gamma_p}}\,
\left[\, 1 + \bar b_1 (p+u)  +\ldots \,\right].
\ee

The free parameters of the models are the residues $d_1^{\rm UV}, d_2^{\rm IR}$ and  $d_3^{\rm IR}$ of the first renormalons and the coefficients $d_0^{\rm PO}, d_1^{\rm PO}$ of the polynomial in (\ref{eq:BBJ}), determined in  \cite{BeJa} as 
\bea 
&& d_1^{\rm UV}=-\,1.56\times 10^{-2},\,\,
d_2^{\rm IR}=3.16,\,\,
d_3^{\rm IR}=-13.5,\nn\\
&& d_0^{\rm PO}=0.781, \,\,
d_1^{\rm PO}=7.66\times 10^{-3},
\eea
by the requirement to reproduce the perturbative coefficients  $c_{n,1}$ in $\overline{{\rm MS}}$ scheme for $n\le 4$, given in (\ref{eq:cn1}), and the estimate $c_{5,1}=283$. 

Once the parameters are fixed,  all the  perturbative coefficients  $c_{n,1}$ for $n>5$ are determined and 
exhibit a factorial increase at high orders.  Their numerical values up to $n=25$ are
\begin{widetext}\bea\label{eq:cn1model}
&& c_{6,1}=3275.45, \,\, c_{7,1}=18758.4, \,\, c_{8,1}=388446, \,\, c_{9,1}=919119, \,\, c_{10,1}= 8.36\times 10^7, \,\,
c_{11,1}= -5.19\times 10^8, \nn\\  
&& c_{12,1}=3.38\times 10^{10},  \,\, 
 c_{13,1}= -6.04\times 10^{11},\,\, c_{14,1}=2.34 \times 10^{13}, \,\, 
 c_{15,1}= -6.52 \times 10^{14},  \,\, ~ c_{16,1}=2.42 \times 10^{16},  \nn\\   
&&c_{17,1}= -8.46\times 10^{17}, \,\, ~ c_{18,1}= 3.36 \times 10^{19}, \,\, c_{19,1}=-1.36 \times 10^{21},   \,\, c_{20,1}=5.92 \times 10^{22},  \,\, c_{21,1}= -2.68 \times 10^{24},  \nn\\ && c_{22,1}=1.28 \times 10^{26}, \,\, c_{23,1}=-6.41 \times 10^{27},  \,\,   c_{24,1}=3.35 \times 10^{29},  \,\,   c_{25,1}=-1.83 \times 10^{31}.\eea
\end{widetext}

A feature of the above model is the relatively large value of the first IR renormalon $d_2^{\rm IR}$. As was argued in \cite{DeMa}, this seems to favor the  FO calculation of $\delta^{(0)}$, while the alternative CI calculation is preferred by situations with a weaker first IR renormalon. Therefore,  alternative models imposing smaller values for the residue of the first IR renormalon,  and even  assuming that this singularity is absent,  have been suggested and investigated in \cite{DeMa, CaFi2011, Abbas:2013usa, BBJ}.  

In our analysis we shall consider for illustration  a model presented in \cite{CaFi2011}, which is defined by the same expressions as in the model \cite{BeJa} for the first three singularities and  the same values of the  residues at $u=-1$ and $u=3$, but a smaller residue at $u=2$, set at $d_2^{\rm IR}=1$.  The model must contain then three additional free parameters  in order to reproduce the first five $c_{n,1}$. In  \cite{CaFi2011},  a quadratic term in the polynomial and two additional IR singularities, at $u=4$ and $u=5$, have been introduced. For convenience, the nature of these additional singularities, which is not known, was assumed to be equal to that of the $u=3$ singularity. Thus, the alternative model is defined by
\bea\label{eq:altBBJ}
&&B_{D, {\rm alt}}(u) =B_1^{\rm UV}(u) +  B_2^{\rm IR}(u) + B_3^{\rm IR}(u)\nn\\&&\hspace{-1.2cm} + \frac{d_4^{\rm IR}}{(4-u)^{3.37}}
+\frac{d_5^{\rm IR}}{(5-u)^{3.37}} + d_0^{\rm PO} + d_1^{\rm PO} u+ d_2^{\rm PO} u^3,
\eea
where the residues of the first renormalons have been fixed at 
\be\label{eq:altdBJ}
d_1^{\rm UV}=-\,1.56\times 10^{-2},\,d_2^{\rm IR}=1,\,
d_3^{\rm IR}=-13.5,
\ee
and the remaining five parameters, determined by matching the coefficients $c_{n,1}$ for $n\le 5$, read
\bea
&&d_0^{\rm PO}=3.2461,\,d_1^{\rm PO}=1.3680, \,d_2^{\rm PO}=0.2785,\nn\\
&& d_4^{\rm IR}=1560.614, \,d_5^{\rm IR}=-1985.73.
\eea

 As above, once the parameters are fixed, the model predicts all the  coefficients  $c_{n,1}$ for $n>5$, which exhibit a factorial increase at large orders. Their numerical values up to $n=25$ are
\begin{widetext}\bea\label{eq:altcn1model}
&&  c_{6,1}=2654.51, \,\, c_{7,1}=7901.76, \,\, c_{8,1}=241607.96, \,\, c_{9,1}=-982236.70, \,\, c_{10,1}= 5.85\times 10^7, \,\,
c_{11,1}= -8.69\times 10^8, \nn\\  
&& c_{12,1}=2.86\times 10^{10},  \,\, 
 c_{13,1}= -6.85\times 10^{11},\,\, c_{14,1}=2.21\times 10^{13}, \,\, 
 c_{15,1}= -6.76\times 10^{14},  \,\, ~ c_{16,1}=2.37\times 10^{16},  \nn\\   
&&c_{17,1}= -8.551\times 10^{17}, \,\, ~ c_{18,1}= 3.34\times 10^{19}, \,\, c_{19,1}=-1.36 \times 10^{21},   \,\, c_{20,1}=5.91 \times 10^{22},  \,\, c_{21,1}= -2.68 \times 10^{24},  \nn\\ && c_{22,1}=1.28 \times 10^{26}, \,\, c_{23,1}=-6.41 \times 10^{27},  \,\,   c_{24,1}=3.35 \times 10^{29},  \,\,   c_{25,1}=-1.83 \times 10^{31}.\eea
\end{widetext}
One can see that at large orders the two models coincide, since the first UV renormalon is modeled in the same way. We emphasize finally that we  consider these models only as a mathematical framework for testing the convergence properties of the various expansions investigated in this work.

\end{document}